\documentclass[12pt]{article}
\pdfoutput=1 %

\usepackage{jheppub} %
\usepackage{graphicx}  %
\usepackage{dcolumn}   %
\usepackage{bm}        %
\usepackage{amssymb}   %
\usepackage{subfigure}
\usepackage{epstopdf}
\usepackage{color}
\usepackage{slashed}
\usepackage[utf8]{inputenc}
\usepackage{multirow}
\usepackage{footnote}
\usepackage{amsmath}
\allowdisplaybreaks[4]
\usepackage{accents}
\newlength{\dhatheight}

\newcommand{\cevns}{{CE$\nu$NS}}
\usepackage{hyperref}
\graphicspath{ {./Figs/} }
\def\figureautorefname~#1\null{Fig.\,#1\null}
\def\tableautorefname~#1\null{Tab.\,#1\null}

\def\equationautorefname~#1\null{Eq.\,(#1)\null}

\title{Sensitivity of the RECODE Reactor CE$\nu$NS Experiment to the Dark Axion Portal}
\author[a,b]{Yingji Shen,}
\author[c]{Jie Tang,}
\author[d]{Li Wang,}
\author[a,b]{Yongcheng Wu,}
\author[e]{Litao Yang}
\affiliation[a]{Department of Physics, Institute of Theoretical Physics and Institute of Physics Frontiers and Interdisciplinary Sciences, Nanjing Normal University, Nanjing, 210023, China}
\affiliation[b]{{Nanjing Key Laboratory of Particle Physics and Astrophysics, Nanjing, 210023, China}}
\affiliation[c]{School of Physics, Jiulonghu Campus, Southeast University, Nanjing 211189, China}
\affiliation[d]{School of Physics and Astronomy, Beijing Normal University, Beijing 100875, China}
\affiliation[e]{Key Laboratory of Particle and Radiation Imaging (Ministry of Education) and Department of Engineering Physics, Tsinghua University, Beijing 100084, China}
\emailAdd{yjshen@njnu.edu.cn}
\emailAdd{tangj@seu.edu.cn}
\emailAdd{wangl@bnu.edu.cn}
\emailAdd{ycwu@njnu.edu.cn}
\emailAdd{yanglt@mail.tsinghua.edu.cn}
\preprint{CPTNP-2026-011}

\abstract{Reactor CE$\nu$NS experiments provide a powerful probe of the physics beyond the Standard Model (BSM) with the intense flux of neutrinos, photons, and othe rparticles produced in nuclear reactors. In this work, we investigate the sensitivity of reactor CE$\nu$NS experiments to the dark axion portal, which connects the axion or axion-like particle to the dark photon. We focus in particular on the RECODE experiment, while also considering other reactor-based experiments such as CONUS and MINER. We find that reactor CE$\nu$NS experiments offer enhanced sensitivity in the sub-MeV region compared with the existing constraints from B factories, and can probe the portal coupling down to $\mathcal{O}(10^{-3})$ for $G_{a\gamma\gamma'}$.
}

\begin{document}
\titlepage
\maketitle

\section{Introduction}

The exploration of light, feebly interacting particles, such as axions, dark photons, and axion-like particles (ALPs), has emerged as a critical frontier in particle physics, offering pathways to address unresolved phenomena like dark matter and the strong CP problem.
In particular, the study of axions originated in solving the strong CP problem~\cite{Peccei:1977hh,Wilczek:1977pj,Weinberg:1977ma,Peccei:1977ur} in the Standard Model (SM). Additionally, both QCD axions and more generalized axion-like particles (ALPs) are considered strong candidates for dark matter~\cite{Duffy:2009ig,Marsh_2016,Battaglieri:2017aum}.
Early studies primarily focused on QCD axions~\cite{Weinberg:1977ma,Peccei:1977hh,Peccei:1977ur,Wilczek:1977pj}, which are pseudoscalar particles capable of resolving the strong CP problem via the dynamical breaking of Peccei-Quinn (PQ) symmetry~\cite{Peccei:1977hh,Peccei:1977ur}.

The study of dark sector particles typically requires extremely high event rates and exceptional detection sensitivity. As a result, nuclear reactors, which serve as intense and well-characterized sources of low-energy neutrinos and photons, provide an ideal environment for exploring new physics beyond the SM (BSM). The idea of utilizing nuclear reactors for axion searches was first proposed by in~\cite{Weinberg:1977ma} and later expanded upon accordingly in~\cite{Donnelly:1978ty}. Many early reactor-based axion experiments placed stringent constraints on the QCD axion model by studying axion attenuation~\cite{Vuilleumier:1981dq,Datar:1982ef,Koch:1986aq,Altmann:1995bw} and decay signatures. Key experimental searches include the $a\to\gamma\gamma$ channel, studied at the Institut Laue-Langevin (ILL) reactor~\cite{Vuilleumier:1981dq} and the Biblis A nuclear power reactor~\cite{Koch:1986aq}, as well as the $a\to e^+e^-$ channel, which was investigated at Bugey Reactor 5~\cite{Altmann:1995bw}. Additionally, searches for two-photon signals from axions produced via neutron capture ($n+p\to d+a$) were conducted at the 500 MW Tarapur light water reactor~\cite{Datar:1982ef}.
To evade experimental constraints, theoretical models propose invisible axions with a symmetry-breaking scale significantly larger than the electroweak scale. These axions are categorized into two primary frameworks: the Kim-Shifman-Vainshtein-Zakharov (KSVZ) model~\cite{Kim:1979if,Shifman:1979if} and the Dine-Fischler-Srednicki-Zhitnitsky (DFSZ) model~\cite{Dine:1981rt}. Experimental searches have explored axion production via neutron capture and nuclear transitions, placing constraints on the axion-photon ($G_{a\gamma\gamma}$) and axion-electron ($G_{aee}$) couplings~\cite{TEXONO:2006spf}.

The introduction of dark photons associated with additional U(1) gauge symmetries opens new avenues for both theoretical and experimental exploration~\cite{Holdom:1985ag,Arkani-Hamed:2008kxc,PhysRevD.80.015003,Goodsell:2011wn,Alexander:2016aln}. In previous studies, the most commonly considered portals connecting the SM to the dark sector include the vector, Higgs, and neutrino portals~\cite{Essig:2013lka,Alexander:2016aln}. The proposal of the dark axion portal~\cite{Kaneta:2016wvf} provides a novel framework that directly links dark photons and axions to the SM. This portal is not a simple combination of the vector and axion portals, but instead represents an independent interaction characterized by the unique coupling $G_{a\gamma\gamma'}$, which connects SM photons, dark photons, and axions. As such, it offers new opportunities for probing dark sector particles~\cite{Kaneta:2017wfh,Choi:2018dqr,deNiverville:2018hrc}. In recent years, this portal has been incorporated into a variety of experimental searches. For instance, B factory experiments such as BABAR~\cite{BaBar:2001yhh} and Belle-II~\cite{Belle-II:2010dht} have placed preliminary constraints on ${\rm MeV}$ to ${\rm GeV}$ scale dark photons through single-photon final states. Reactor experiments, including RENO and NEOS~\cite{NEOS:2016wee,Proceedings:2019mnq,RENO:2010vlj}, have explored lower mass regimes by exploiting intense photon fluxes, while fixed target neutrino experiments (FTNEs) such as LSND~\cite{LSND:1996jxj} and MiniBooNE~\cite{MiniBooNE:2013dds}, together with beam dump experiments like CHARM~\cite{Gninenko:2011uv}, provide complementary sensitivity in the intermediate mass range. With continued improvements in experimental precision and the development of joint analyses involving multiple portals, studies of the dark axion portal are expected to play an increasingly important role in elucidating the structure and interactions of the dark sector. In particular, reactor \cevns{}{} experiments offer a unique opportunity to probe the dark axion portal by studying the production and decay of axions and dark photons in the reactor environment. Compared to FTNEs, the dark photons and axions produced in reactor experiments and their signals can be up to $10^6$ times more abundant than those produced in FTNEs. So we investigate the dark axion portal in reactor neutrino experiments. In this study our analysis focuses on the RECODE experiment, along with other reactor experiments such as MINER, and CONUS extending the coverage from a specific region $m_a\ll m_{\gamma'}$ to the full mass parameter space. We perform a numerical study of axion and dark photon production from the reactor and detection through both decay and scattering processes mediated by the dark axion portal and derive constraints on the relevant parameter space. These results complement existing constraints, enable systematic comparisons among different experimental probes, and extend the accessible parameter space for future studies.

The paper is organized as follows. \autoref{sec:Reactor CEvNS experiment} briefly discuss the reactor \cevns{} experiments, highlighting the key configurations of the RECODE experiment considered here. \autoref{sec:Model} outlines the theoretical framework of the dark axion portal and details the production and detection of axions and dark photons. The sensitivities of \cevns{} experiments are also estimated. \autoref{sec:Conclusion} summarizes the main results.

\section{Reactor \cevns{} experiment}
\label{sec:Reactor CEvNS experiment}
Coherent elastic neutrino-nucleus scattering (\cevns{}) is a neutral-current process within the SM in which a neutrino interacts with the target nucleus as as whole and is first predicted in 1974~\cite{Freedman:1973yd}. The process is characterized by a cross-section that is enhanced approximately by the square of the number of neutrons ($\propto N^2$) in the target nucleus. Despite this enhancement, the experimental observation of \cevns{} remained elusive for over four decages due to the technically challenging signature of such process: the extremely low nuclear recoil energy which is typically in the keV to sub-keV range. The development in the detection technology of the dark matter direct detection makes such detection possible.

The importance of \cevns{} extends across particle physics, nuclear physics, and astrophysics~\cite{Abdullah:2022zue,AristizabalSierra:2019ykk}. Within the SM, it offers a distinct channel for precision tests of the electroweak parameters, especially the weak mixing angle ($\sin^2\theta_W$) at low scale~\cite{ParticleDataGroup:2018ovx,Safronova:2017xyt,Cadeddu:2021dqx,Majumdar:2022nby}. \cevns{} data also allows for the extraction of the nuclear form factors and the neutron radius, providing insights into the neutron distribution inside the nuclei. Furthermore, the process plays a critical role in supernova dynamics and constitutes an irreducible neutrino floor/fog background for the next-generation dark matter searches~\cite{Abdullah:2022zue,XENON:2024hup,DeRomeri:2025nkx,Billard:2013qya}.

Since the first observation of \cevns{} process by the COHERENT collaboration in 2017~\cite{COHERENT:2017ipa}, lots of efforts have been put into the \cevns{} detection across different kinds of detectors and neutrino sources. Utilizing the high-intensity stopped pion decay neutrino source at the Spallation Neutron Source (SNS) at Ork Ridge National Laboratory, COHERENT detected \cevns{} on a CsI[Na] scintillator~\cite{COHERENT:2017ipa,COHERENT:2021xmm}. This was subsequently followed by the detection on liquid argon (LAr)~\cite{COHERENT:2020iec}, and more recently germanium targets~\cite{COHERENT:2026ewu}. Beyond the accelerator-based sources, the field has recently expanded to solar neutrinos, with XENONnT and PandaX-4T dark matter experiments reporting the first evidence of \cevns{} induced by solar neutrinos~\cite{XENON:2022ltv,XENON:2024ijk,PandaX:2024muv,PandaX:2024cic}.

Another major part in the measurement of \cevns{} process is the detection using reactor antineutrinos. Nuclear reactors provide the most intense fluxes of low energy electron antineutrinos, offering a complementary sources to the SNS. However, the lower neutrino energy results in even smaller nuclear recoil energy, and thus requires ultra-low threshold detectors and stringent background controls. While it is still challenging, the CONUS+ experiment recently reported the first evidence of observation of the reactor \cevns{} process using germanium (Ge) detectors~\cite{CONUS:2024lnu}. Other experimental efforts, such as the Dresden-II~\cite{Colaresi:2022obx}, RED-100~\cite{Akimov:2017hee}, NUCLEUS~\cite{Strauss2017}, MINER~\cite{MINER:2016igy,James_2019}, CONNIE~\cite{CONNIE:2016ggr}, RELICS~\cite{RELICS:2024opj}, RECODE~\cite{Yang:2024exl}, Ricochet~\cite{Billard:2016giu}, CONUS~\cite{Buck:2020opf}, NEOS~\cite{NEOS:2016wee,Proceedings:2019mnq,RENO:2010vlj}, TEXONO~\cite{Wong:2015kgl} continues to advance the measurement near the reactors.

The reactor-based \cevns{} experimetns are also very unique for probing physics beyond the SM. The low enrgy of the neutrino flux from the reactor although challenge the detector, also results in that the interaction between the neutrino and the nucleus happens in the fully coherent regime, where the nuclear form factor is approximately 1 $F(Q^2)\approx 1$ which minimizes the theoretical uncertainty related to the nuclear structure, allowing any deviation in the measurement to be attributed to potential new physics. Hence, reactor \cevns{} measurement provide powerful constraints on non-standard neutrino interactions (NSI)~\cite{DeRomeri:2022twg,Coloma:2023ixt,Khan:2019cvi,Coloma:2022avw}, sterile neutrinos~\cite{Giunti:2019aiy,PhysRevD.86.013004,PhysRevD.96.063013,PhysRevD.101.075051}, and electromagnetic properties of the neutrino, including the magnetic moment, charge radius~\cite{Giunti:2022aea,AtzoriCorona:2022qrf,Khan:2019cvi,Coloma:2022avw}. Further, the \cevns{} process will also receive contributions from possible light, weakly coupled new particles~\cite{Khan:2019cvi,AtzoriCorona:2022moj,Cadeddu:2020nbr}. The precise measurement can put strong constraints on light vector or scalar mediators~\cite{Khan:2019cvi,Coloma:2022avw,AtzoriCorona:2022moj,Cadeddu:2020nbr}. On the other hand, the reactor core also produce intense low energy photon flux~\cite{bechteler_faissner_yogeshwar_seyfarth_1984}, which can produce light BSM particles through the interactions with the materials within the reactor. In this context, the axion-like-partile (ALP) has been considered with its interactions with photon and electron near the reactors~\cite{Park:2017prx,Ge:2017mcq,AristizabalSierra:2020rom,Deniverville:2020rbv}. The \cevns{} experiment near reactor can be complementary to other measurement around MeV scale, in particular can cover the ``cosmological triangle'' region.

\begin{table}[!bp]
\centering
\small
\resizebox{\textwidth}{!}{
\begin{tabular}{|c|c|c|c|c|c|c|c|}
\hline\hline
  Experiment &\begin{tabular}{c} Detector \\ materials\end{tabular}& \begin{tabular}{c} Detector \\  mass/volume \end{tabular} & \begin{tabular}{c} Reactor \\ Power \end{tabular} & \begin{tabular}{c} Detector \\ distance \end{tabular} & \begin{tabular}{c} Background \\ Rate\end{tabular} & \begin{tabular}{c} Energy \\ Threshold \end{tabular}& Reference \\
\hline
RECODE & Ge&10 \rm kg/847.8 {$\rm cm^3$} & 3.4 {\rm GW} & \begin{tabular}{c} 11 {\rm m}(near) \\ 22 {\rm m}(far) \end{tabular} & 2 {\rm cpkkd} & 160 {\rm eV} & \cite{Yang:2024exl} \\ %
\hline
CONUS & Ge& 4\rm kg/751.46 {$\rm cm^3$} & 3.9 {\rm GW} & 17 {\rm m} & 100 {\rm cpkkd} & 300 {\rm eV}  & \cite{CONUS:2024lnu} \\ %
\hline
MINER &Ge/Si& 20\rm kg/3085.2 {$\rm cm^3$} & 1 {\rm MW} & 2.835 {\rm m} & 10 {\rm cpkkd} & 100 {\rm eV}  & \cite{MINER:2016igy} \\ %
\hline\hline
\end{tabular}
}
\caption{The configurations of several reactor-based \cevns{} experiments. }
\label{tab:reactor_CEvNS_experimetns}
\end{table}

In this paper, we will focus on the RECODE experiment~\cite{Yang:2024exl} and estimate its potential to probe the dark axion portal which adds dark photon and ALP to the SM content with their interaction with photon. The RECODE experiment focuses on the detection of \cevns{} event induced by reactor neutrino using high purity germanium detector near the Sanmen nuclear power plant with a thermal power of 3.4 GW. The detector contains two arraies, each contains about 5 kg cylindrical high-purity germanium. The two arraries will be placed at distances of about 11 m and 22 m respectively from the reactor core. As a comparison, the sensitivity of MINER and CONUS experiments will also be considered. We listed the most important information about these experiments in~\autoref{tab:reactor_CEvNS_experimetns}.

\section{The Dark Axion Portal}
\label{sec:Model}

In this work, we consider the SM extended with ALP ($a$) and dark photon ($\gamma'$) with the following extra interactions:
\begin{align}
    \mathcal{L}_{\rm DAP}\supset \frac{G_{a\gamma\gamma'}}{2}aF_{\mu\nu}\tilde{Z}^{\prime\mu\nu},
\end{align}
where $Z'_{\mu\nu}$ and $F_{\mu\nu}$ are the field strength of the dark photon $\gamma'$ and SM photon $\gamma$, respectively. Such coupling can be induced if there exists fermions at high energy scale that are charged under $U(1)_{PQ}$, $U(1)'$ as well as the electromagnetic gauge group~\cite{KunioKaneta_2016}. Note that, through the same mechanism, the ALP can couple to two dark photon as well. Further, the ALP and dark photon can also have their own interactions with the SM particles. However, in this study, we focus on the interplay between the ALP and dark photon through the above portal coupling, and hence assuming all other couplings are small.

\subsection{ALP and Dark Photon Production from Reactors}
\label{sec:productions}

The nuclear reactors, besides providing clean energy, also produce the largest artificial neutrino flux and hence play a crucial role in scientific studies, especially for the neutrino physics. However, on the other hand, the reactor core also produce huge amount of low energy photons, of which the spectrum can be well approximated by~\cite{Bechteler_1984}:
\begin{align}
    \frac{dR_\gamma}{dE_\gamma} = \frac{0.58\times 10^{21}}{\rm sec\cdot MeV}\frac{P}{\rm GW}e^{-1.1\frac{E_\gamma}{\rm MeV}},
\end{align}
where $P$ is the thermal power of the reactor, $E_\gamma$ is the photon energy. Such intense low energy photon flux has been utilized in recent years for phenomenological studies of dark photons, ALPs and many other dark sector particles~\cite{Aristizabal_2020,HyangKyu_2017,Patrick_2020,Shao-Feng_2017}.

\begin{figure}[!btp]
    \centering
    \includegraphics[width=0.8\linewidth]{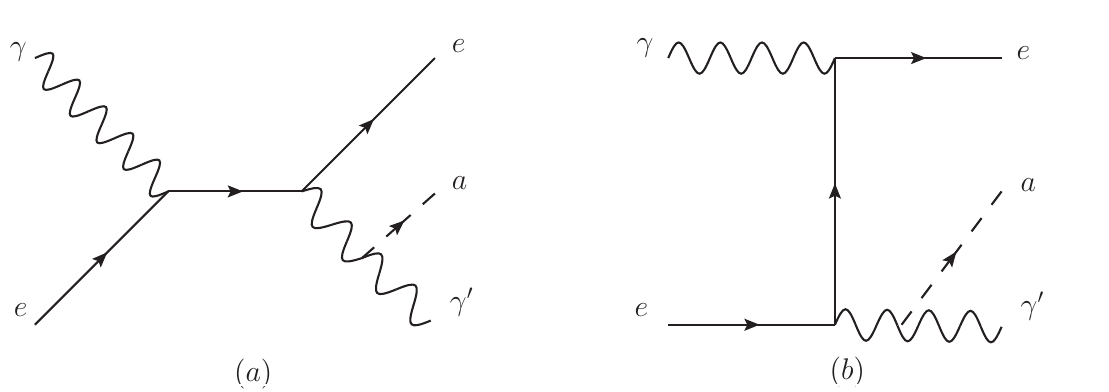}
    \caption{Feynman diagrams for the process $\gamma e^-\to\gamma' ae^-$ with the dark axion portal $G_{a\gamma\gamma'}$.}
    \label{fig:Feynman_Gagammagamma'}
\end{figure}

Within dark axion portal, the ALP and dark photon can be produced simultaneously through the scattering of the low energy photon with materials $\gamma e\to e\gamma' a$ controled by the portal coupling $G_{a\gamma\gamma'}$. The relevant Feynman diagrams are shown in~\autoref{fig:Feynman_Gagammagamma'}. Note that, the ALP and dark photon can be produced independently through many other processes. However, in this study, we will focus on this process through the dark axion portal coupling with the dark photon and ALP produced simultaneously.

\begin{figure}[!bp]
    \centering
    \includegraphics[width=0.6\linewidth]{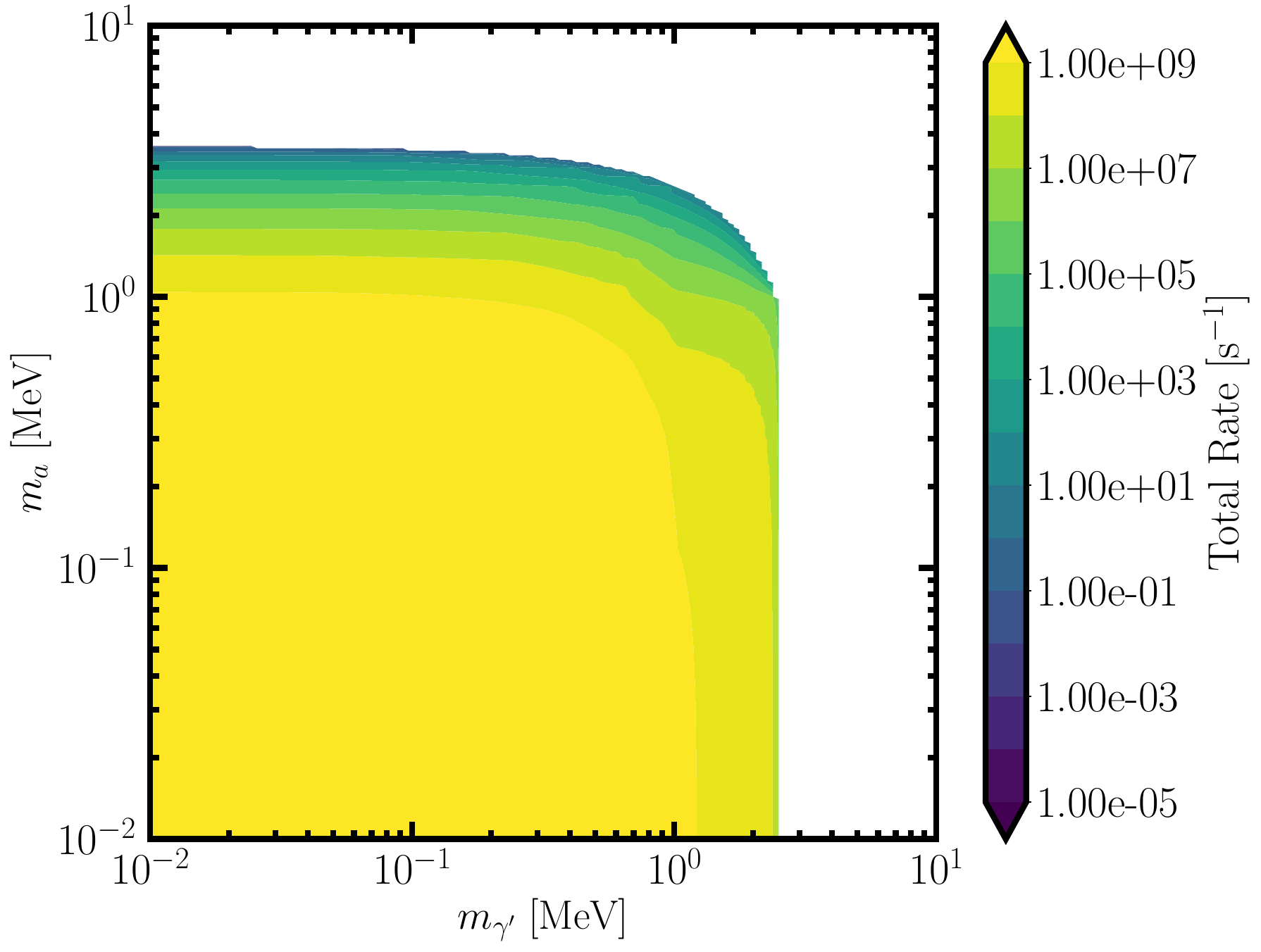}
    \caption{The cross section of $\gamma e \to e \gamma' a$ intergrating over the energy of photons from reactors with $G_{a\gamma\gamma'}=1\,{\rm GeV^{-1}}$.}
    \label{fig:production_cs}
\end{figure}

\begin{figure}[!tbp]
    \centering
    \includegraphics[width=0.48\textwidth]{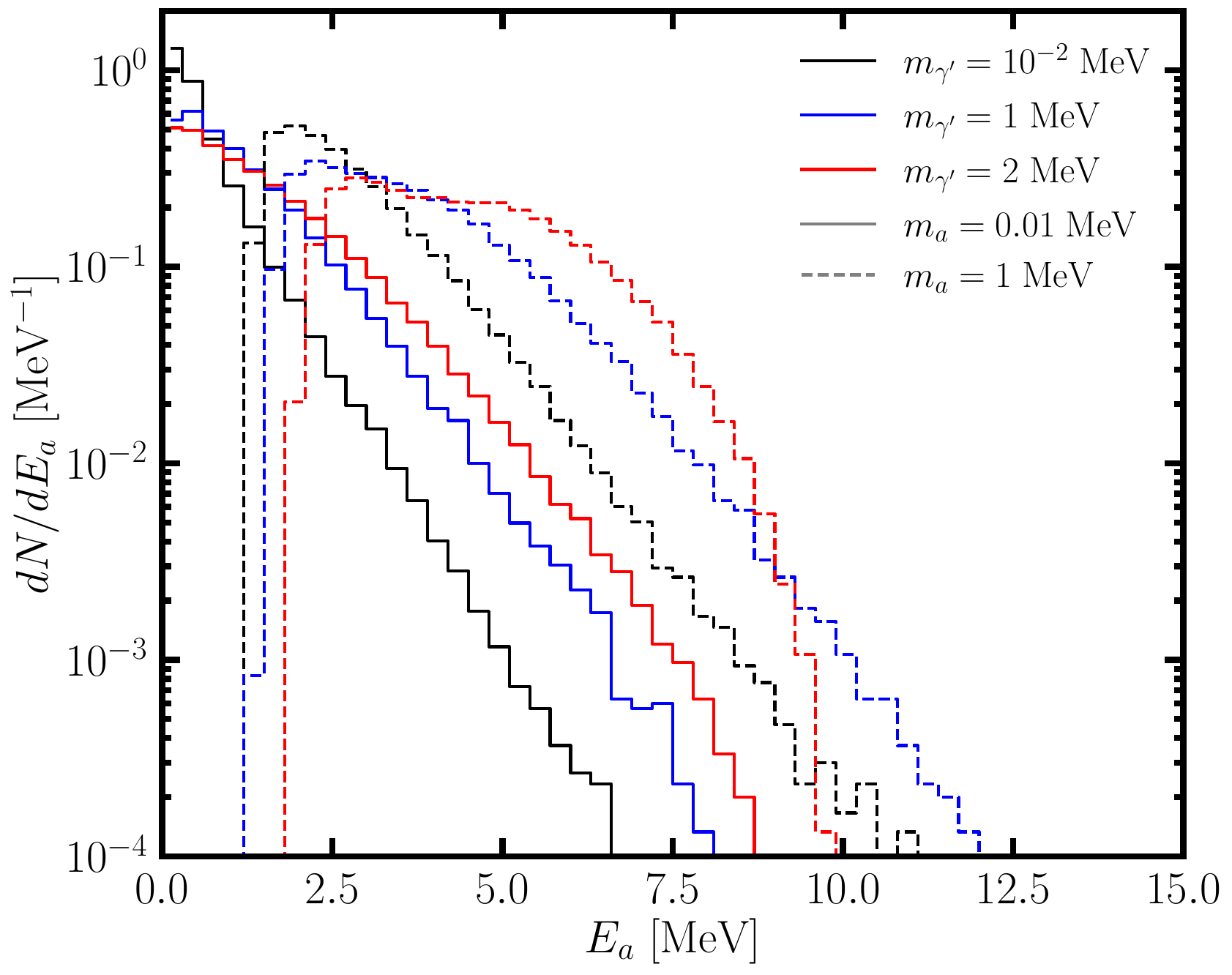}
    \includegraphics[width=0.48\textwidth]{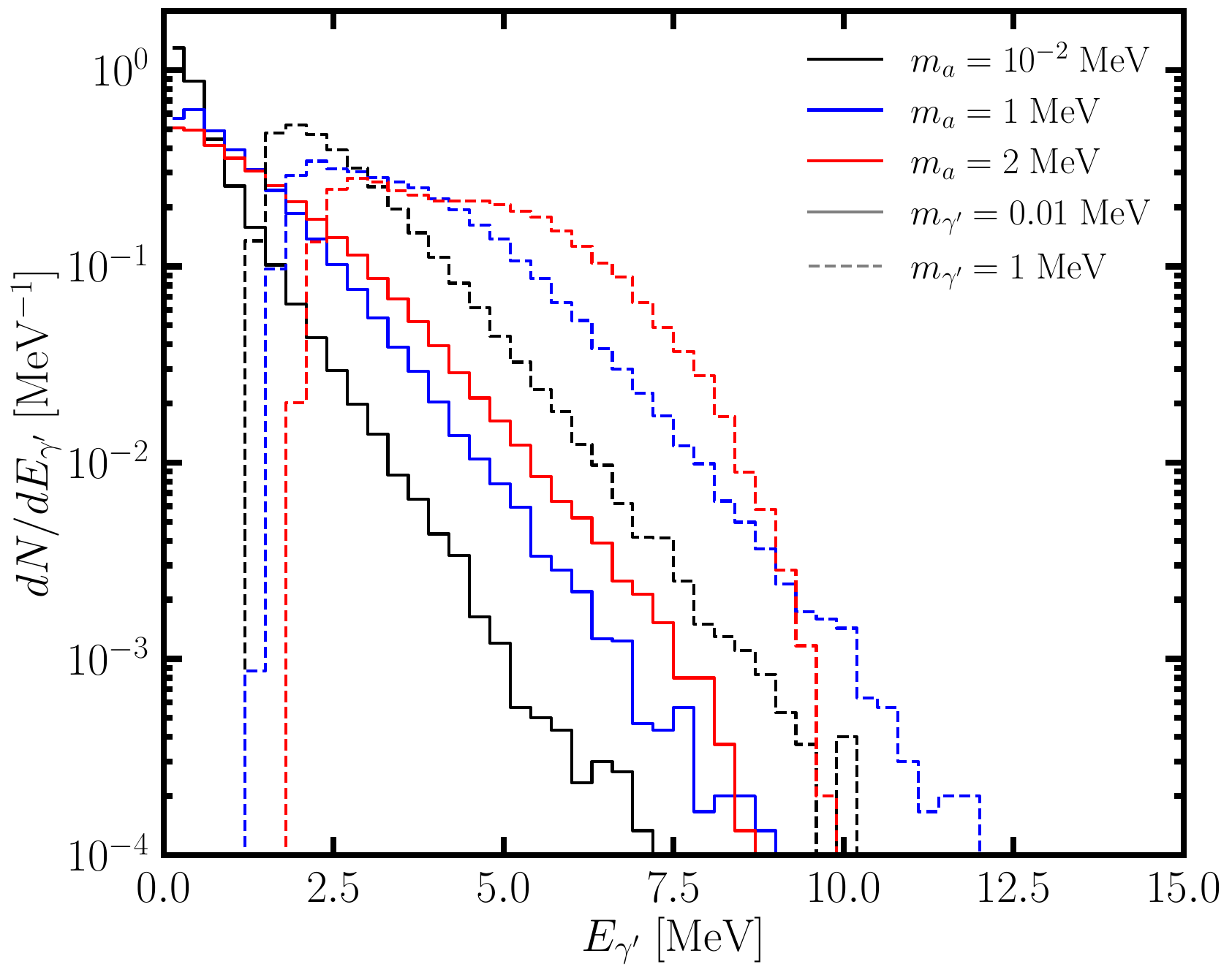}
      \caption{The normalized energy spectra for ALP (left) and dark photon (right) produced through $\gamma e^-\to e^- \gamma' a$ process from reactor for different choices of the masses.}
    \label{fig:spectrum_ALP_DP}
\end{figure}

For either ALP and dark photon, the spectrum produced from reactor core can be obtained as
\begin{align}
    \frac{dR_\varphi}{dE_\varphi} = \int dE_\gamma \frac{1}{\sigma_{\rm tot}} \frac{d\sigma_{\gamma e\to e\gamma' a}}{dE_\varphi}\frac{dR_\gamma}{dE_\gamma}
\end{align}
where $\varphi$ can be either ALP $a$ or the dark photon $\gamma'$. $\sigma_{tot}$ is the total interaction cross section between photons and matter which is obtained from~\cite{HyangKyu_2017,XCOM}.
The total production cross sections integrating over the energy of photons from the reactor, and the final state energy in $m_{\gamma'}$-$m_a$ plane with $G_{a\gamma\gamma'}=1\, {\rm GeV^{-1}}$ for $\gamma e\to e\gamma'a$ is shown in~\autoref{fig:production_cs}. Since the photon energy from the reactor is limited to several MeV, the masses of the ALP and dark photon that can be produced are also restricted to at most the MeV scale. When the mass is close to such threshold, the cross section is extremely small. However, when the mass is below MeV, compared with the photon energy from the reactor, they are effectively massless, the cross section is almost independent of the masses and is sufficient for us to make detection.
The normalized energy spectra are also shown with various different choice of the masses for ALP and dark photon in~\autoref{fig:spectrum_ALP_DP}. It is clear that for a given mass of ALP (dark photon), increase the mass of dark photon (ALP) will shift the spectrum towards the high energy end.

\subsection{ALP and Dark Photon Detection Near Reactor}
\label{sec:Results}

The ALP and Dark Photon produced from the reactor core as discussed above can escape the shelding of the reactor due to their weak interactions with the SM particles. These particles can be detected outside the reactor mainly through either their decays within the detector or scattering with the nucleons within the detector. Similar to the production, for the detection, we will also only focus on $G_{a\gamma\gamma'}$ coupling.

\subsubsection{Detection through the Decay}
In our setup, only the heavier one can be detected through the corresponding decays, $a\to\gamma\gamma'$ for the ALP or $\gamma'\to\gamma a$ for the dark photon. The corresponding decay event rate is given as
\begin{align}
    \frac{dN_\varphi^D}{dE_\gamma} = T \frac{\mathcal{A}}{4\pi L^2}\int dE_\varphi\frac{dR_\varphi}{dE_\varphi}\mathcal{P}_{\rm decay}\frac{d\mathcal{P}_{E_\varphi\to E_\gamma}}{dE_\gamma},
\end{align}
where $T$ is the operation time of the experiment, $\mathcal{A}$ is the transverse area of the detector, $L$ is the distance between the detector and the reactor core. $\frac{d\mathcal{P}_{E_\varphi\to E_\gamma}}{dE_\gamma}$ is the differential probability of obtaining a photon with $E_\gamma$ from decaying particle $\varphi$ with energy $E_\varphi$ which is given by
\begin{align}
    \frac{d\mathcal{P}_{E_\varphi\to E_\gamma}}{dE_\gamma} = \frac{1}{2\beta\gamma E_{\gamma}^{\rm cm}}\Theta(1-|c_\theta|),
\end{align}
where $\Theta(\cdot)$ is the step function, $\beta/\gamma$ are the lorentz factors of the decaying particle and $E_\gamma^{\rm cm}=\frac{m_0^2-m_1^2}{2m_0}$ is the photon energy in the rest frame of the decaying particle with $m_0$ the mass of the decaying particle and $m_1$ the mass of the other decay product, $c_\theta$ is given as
\begin{align}
    c_\theta = \frac{E_\gamma-\gamma E_\gamma^{\rm cm}}{\beta\gamma E_\gamma^{\rm cm}}.
\end{align}
Further $\mathcal{P}_{\rm decay}$ indicates the probability of $\varphi$ decaying within the detector and is given by
\begin{figure}[!bp]
    \centering
    \includegraphics[width=0.48\textwidth]{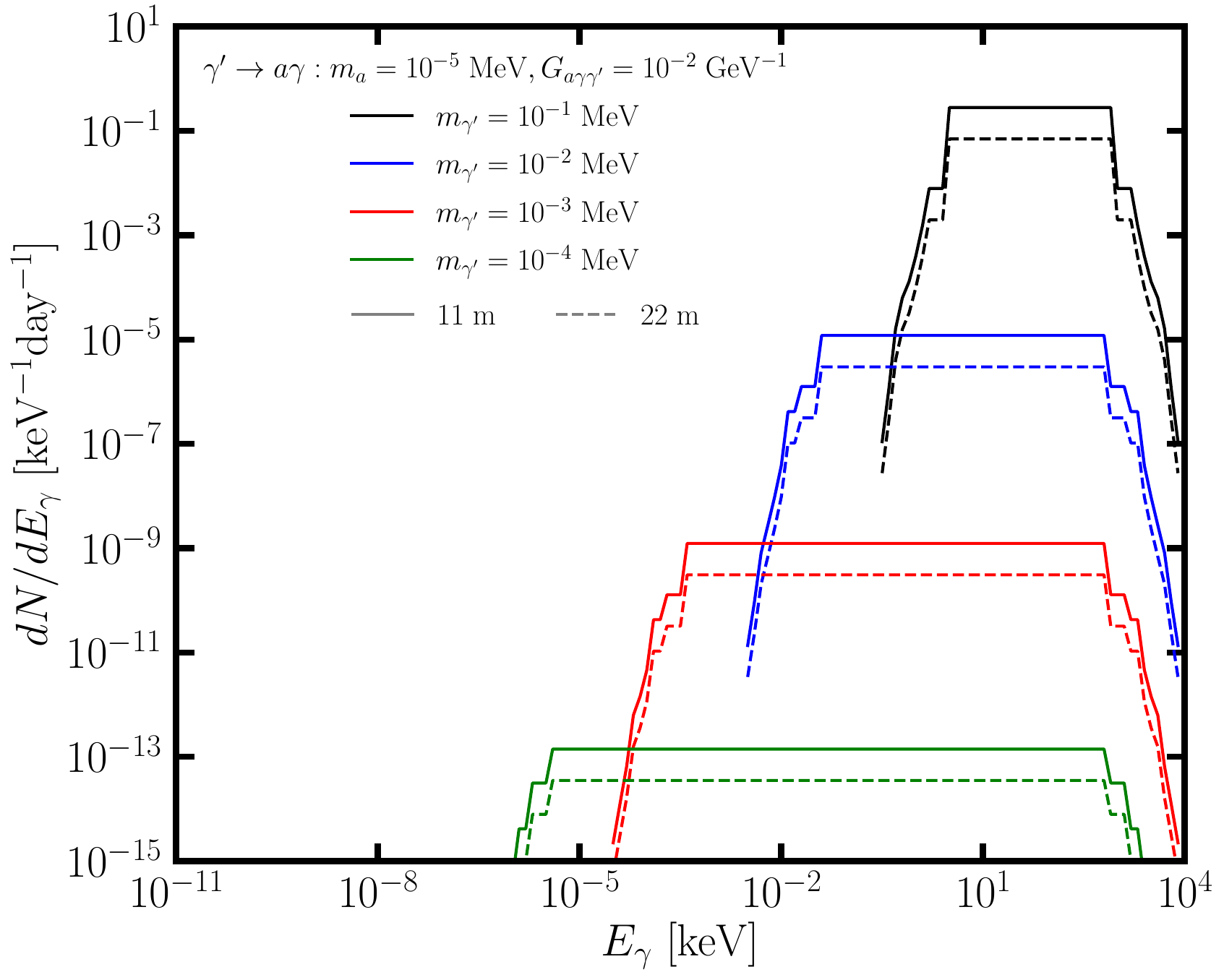}
    \includegraphics[width=0.48\textwidth]{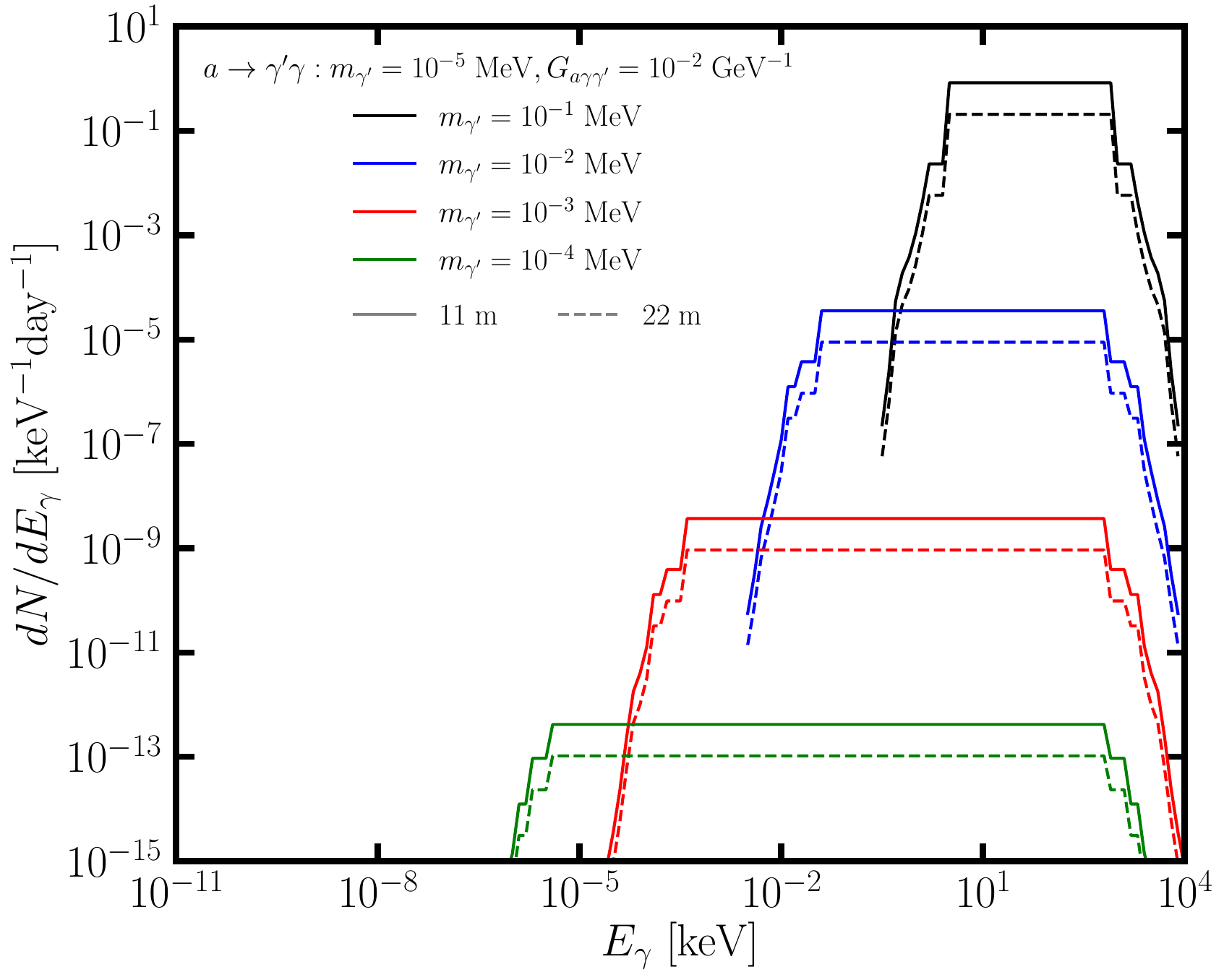}
    \caption{The event spectrum from ALP (left) and dark photon (right) decay for detectors located at 11 m (solid lines) and 22 m (dashed lines) from the reactor with $G_{a\gamma\gamma'}=10^{-2}\,\rm GeV^{-1}$.}
    \label{fig:decay_spectrum_ALP_DP}
\end{figure}
\begin{align}
    \mathcal{P}_{\rm decay} &= \exp\left(-\frac{L}{\beta\gamma c\tau}\right)-\exp\left(-\frac{L+\delta L}{\beta\gamma c\tau}\right)\nonumber \\
    &\approx \exp\left(-\frac{L}{\beta\gamma c\tau}\right)\frac{\delta L}{\beta\gamma c\tau}
\end{align}
where $\delta L$ is the length of the detector, $c\tau$ is the decay length of the decaying particle. The last approximation is valid for a relatively small detector (compared with the decay length). With this approximation, the event rate from ALP/dark photon decay is given as
\begin{align}
    \frac{dN_\varphi^D}{dE_\gamma} = T \frac{\mathcal{V}}{4\pi L^2}\int dE_\varphi\frac{dR_\varphi}{dE_\varphi}\frac{\exp\left(-\frac{L}{\beta\gamma c\tau}\right)}{\beta\gamma c\tau}\frac{d\mathcal{P}_{E_\varphi\to E_\gamma}}{dE_\gamma}
\end{align}
which is now explicitly proportional to the volume of the detector $\mathcal{V}$. The thus obtained event rate spectra from dark photon and ALP decay are presented in~\autoref{fig:decay_spectrum_ALP_DP} for various choices of the masses with $G_{a\gamma\gamma'}=10^{-2}\,\rm GeV^{-1}$.

\begin{figure}[!bp]
    \centering
    \includegraphics[width=0.48\linewidth]{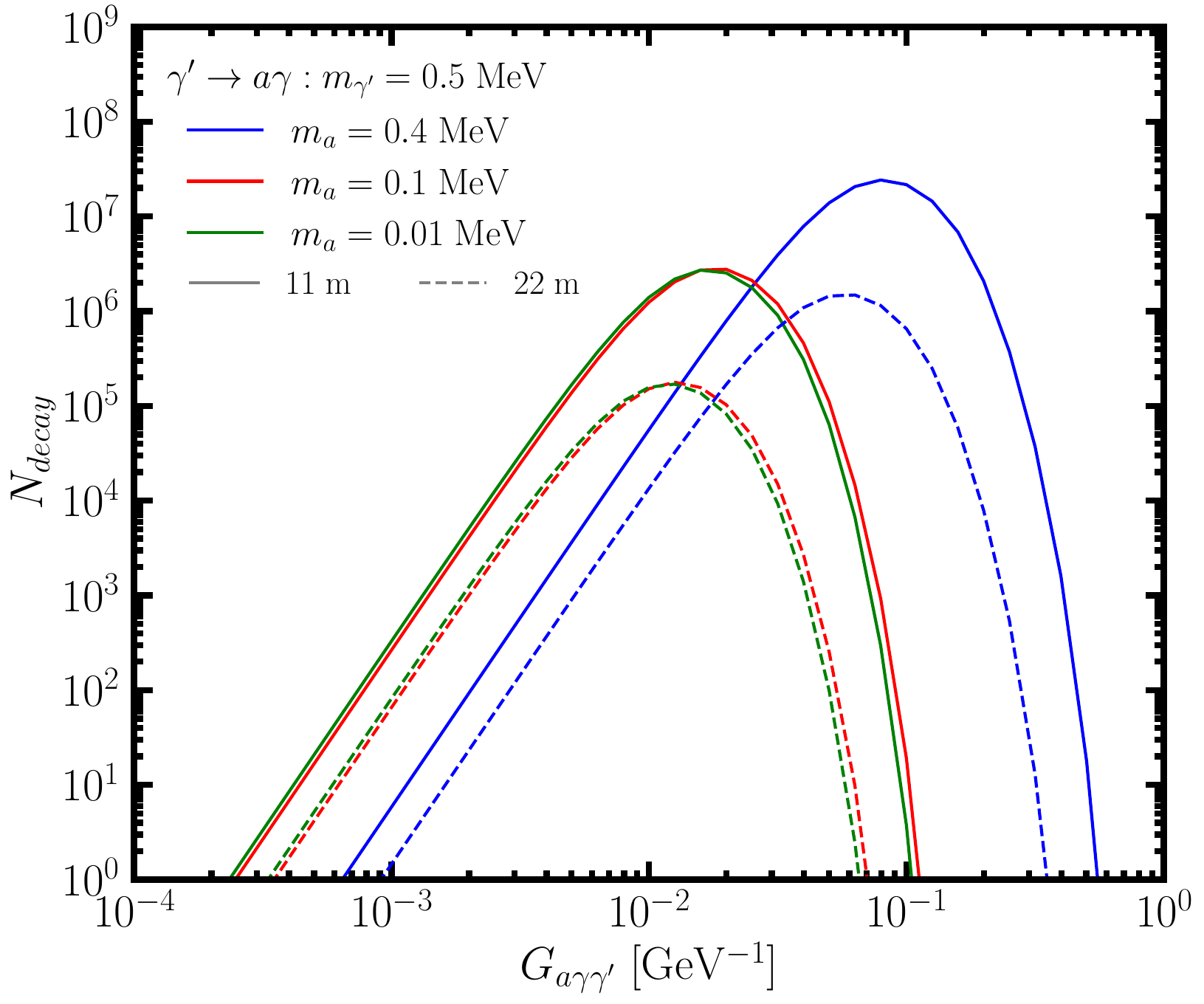}
    \includegraphics[width=0.48\linewidth]{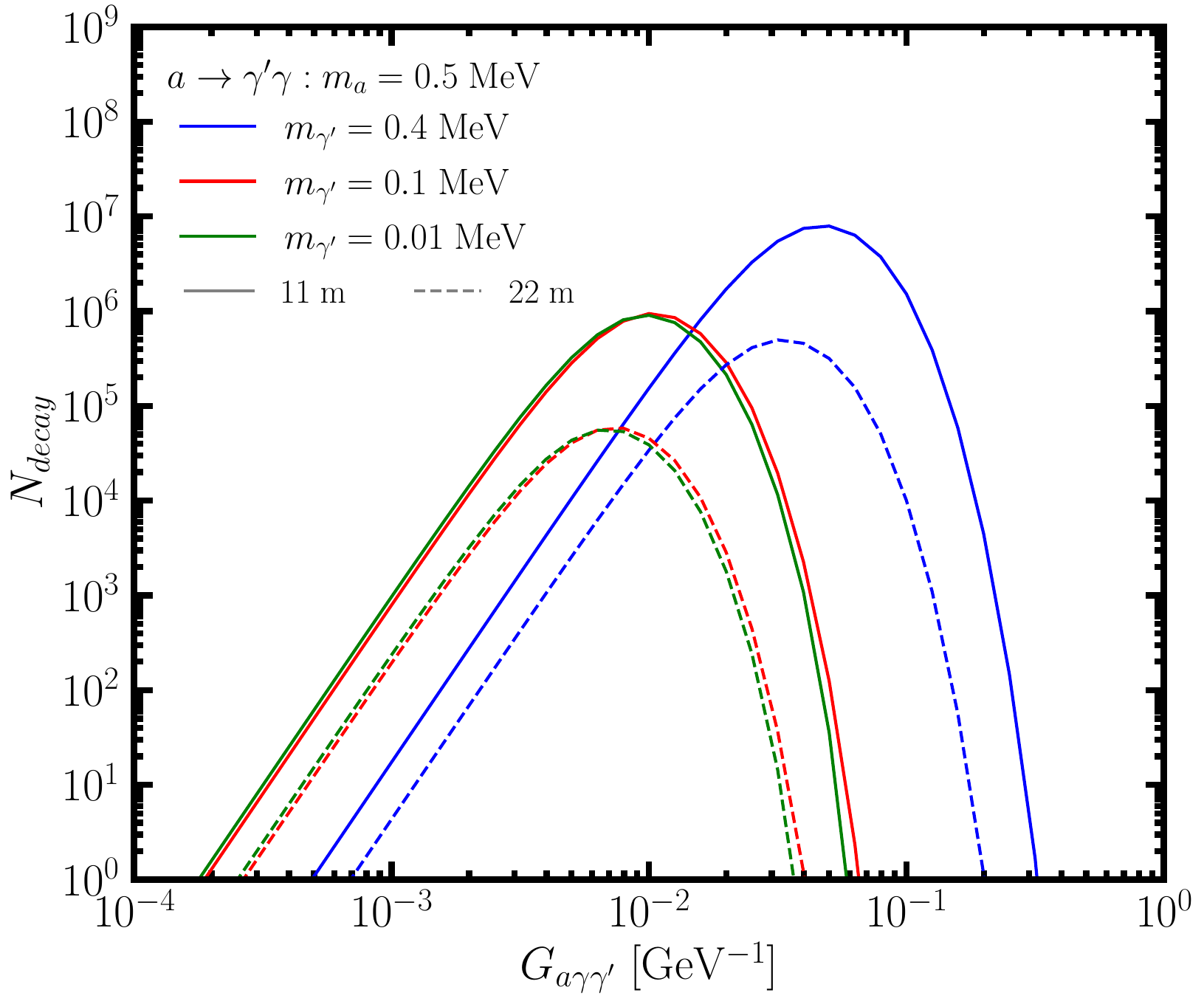}
    \caption{The event rate per year from dark photon (left panel) and ALP (right panel) decays as function of the coupling $G_{a\gamma\gamma'}$ for several difference benchmarks of dark photon and ALP masses.}
    \label{fig:rate_from_decay}
\end{figure}

Integrating over the photon energy above the detector threshold, we obtain the event with one year operation from ALP/dark photon decays as function of the coupling $G_{a\gamma\gamma'}$ showing in~\autoref{fig:rate_from_decay}. It is clear that with given mass benchmark, the event rate peaks at specific value of $G_{a\gamma\gamma'}$. Larger coupling leads to shorter life time, the dark photon or ALP cannot reach the detectors. On the other hand, smaller coupling will reduce the production rate of both dark photon and ALP, and leads to longer life time that may easily bypass the detector without decaying.

\begin{figure}[!tbp]
\centering
\includegraphics[width=0.48\textwidth]{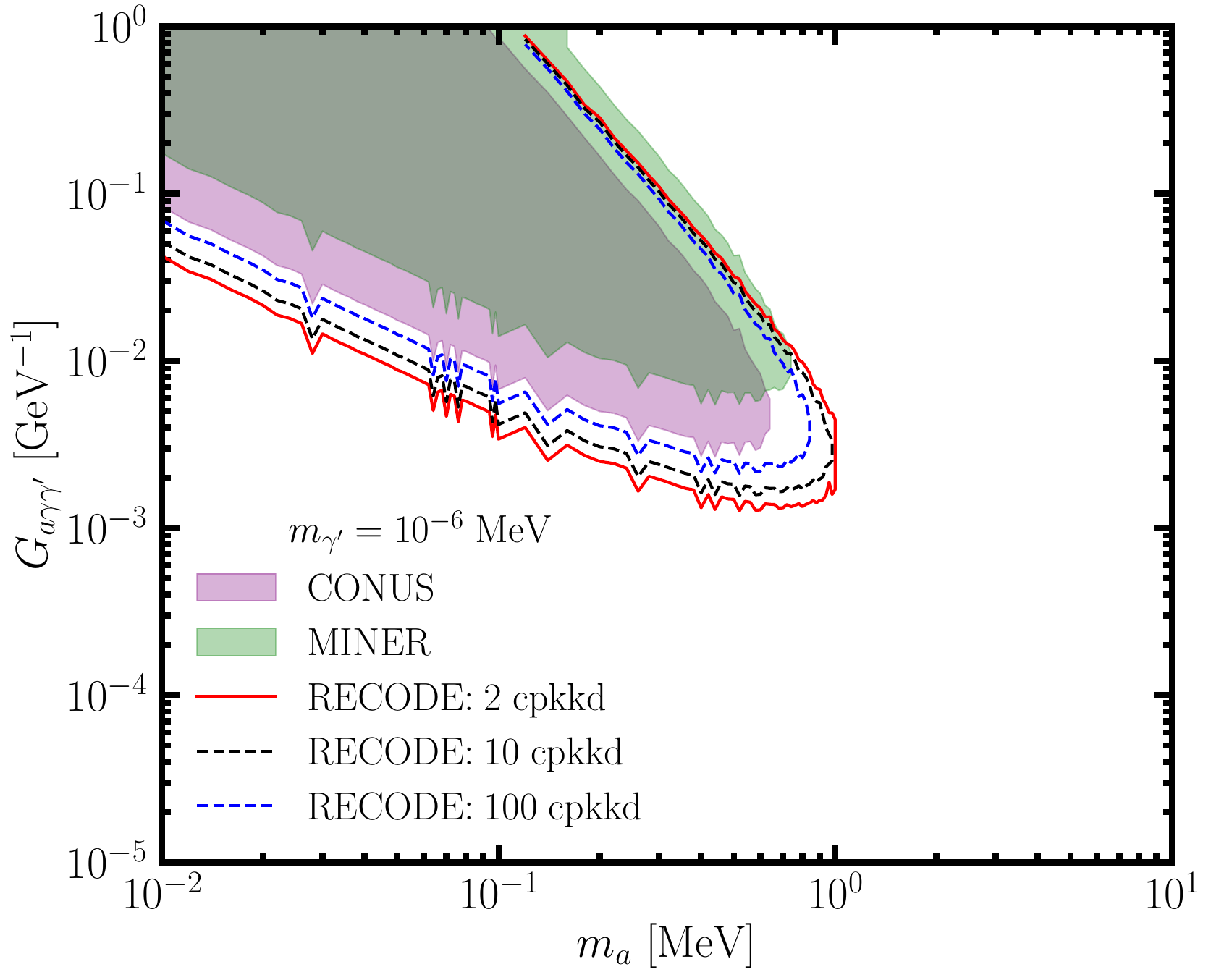}
\includegraphics[width=0.48\textwidth]{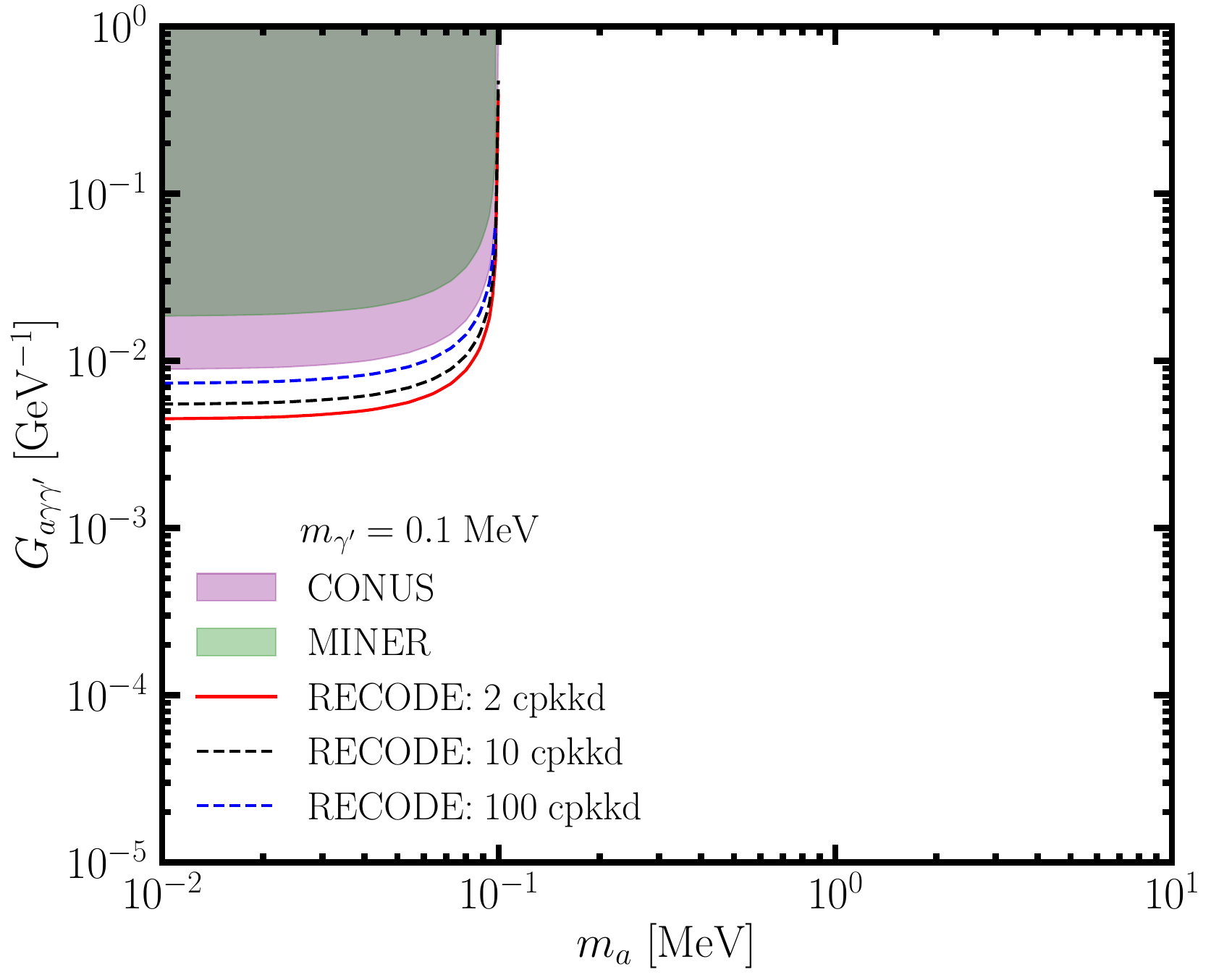}\\
\includegraphics[width=0.48\textwidth]{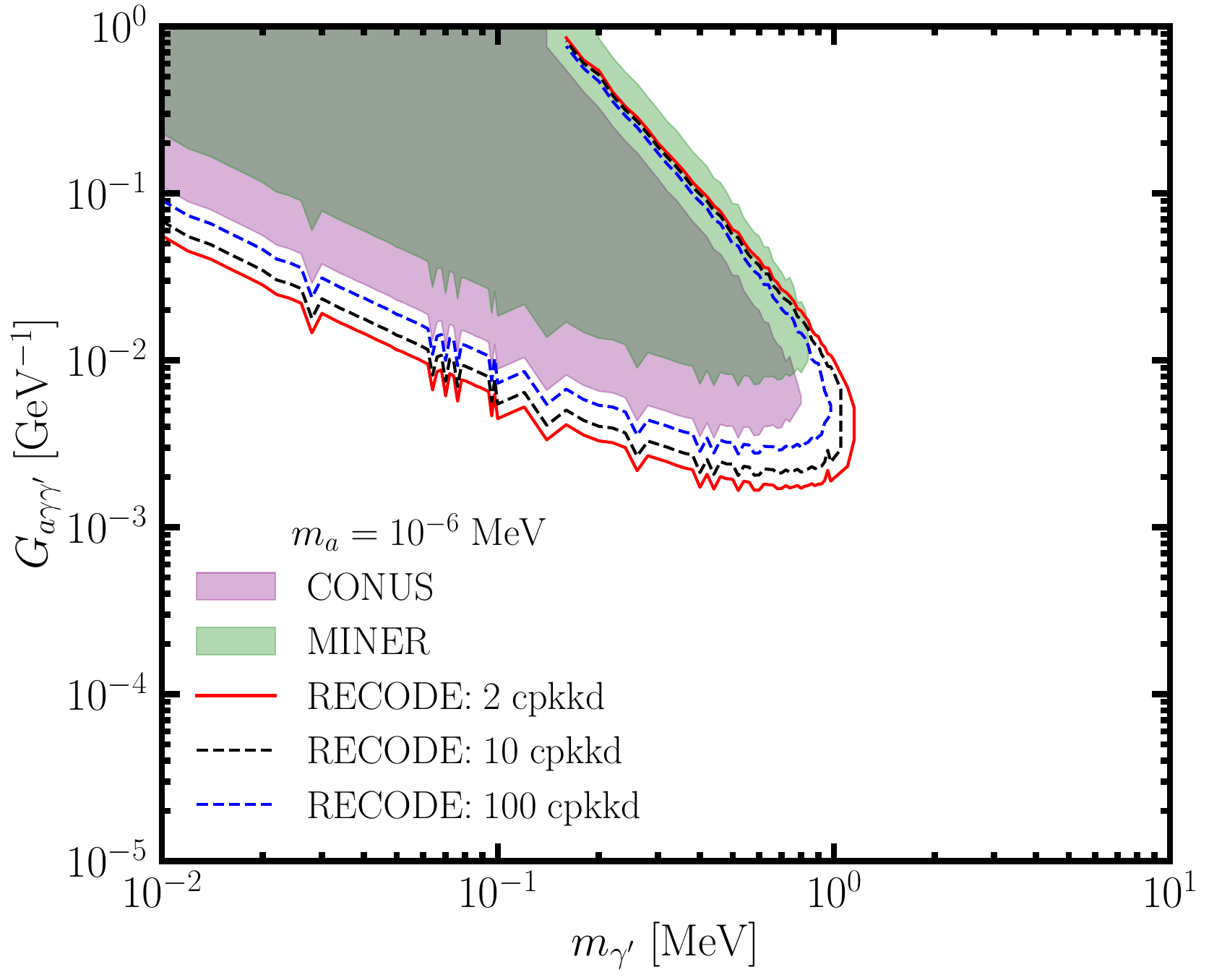}
\includegraphics[width=0.48\textwidth]{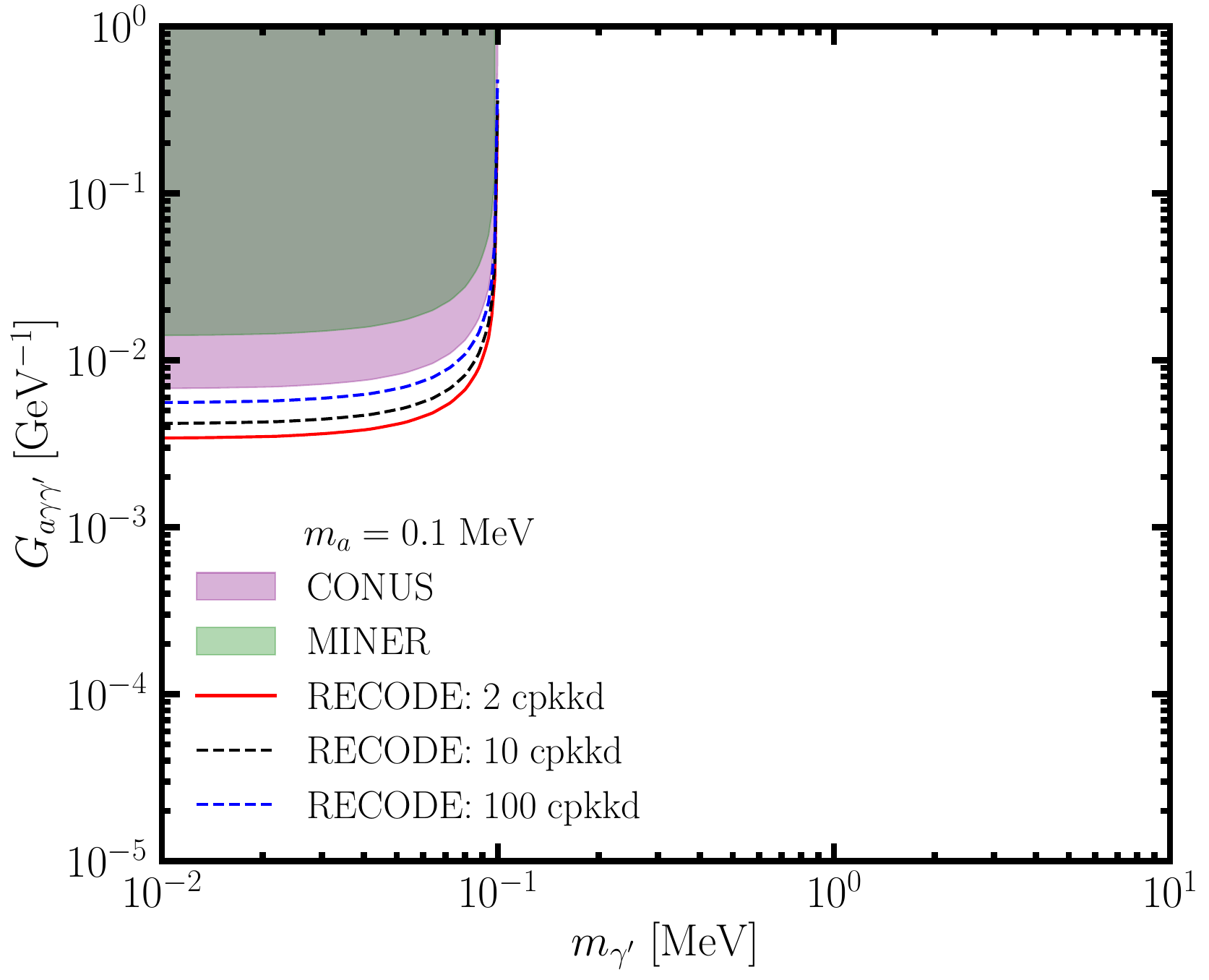}
\caption{The projected 95\% C.L. exclusion region of the RECODE, CONUS, and MINER experiments with one year operation obtained from decay events. The sensitivities are given in $m_a$-$G_{a\gamma\gamma'}$ (upper panels) and $m_{\gamma'}$-$G_{a\gamma\gamma'}$ (lower panels) planes. The other mass ($m_{\gamma'}$ for upper panels, $m_a$ for lower panels) is fixed at $10^{-6}\,\rm MeV$ (left panels) and $0.1\,\rm MeV$ (right panels).}
\label{fig:Ndecay_Ga_ma_dark_axion_portal}
\end{figure}

Combining the event rate from dark photon and ALP decay, considering the experiment setups given in~\autoref{tab:reactor_CEvNS_experimetns}, we obtain the exclusion region at 95\% C.L. for one-year operation with one of the masses fixed as shown in~\autoref{fig:Ndecay_Ga_ma_dark_axion_portal} in both $m_a$-$G_{a\gamma\gamma'}$ plane (upper panels) with $m_{\gamma'}$ fixed at $10^{-6}$ (left) and $0.1$ (right) MeV and $m_\gamma'$-$G_{a\gamma\gamma'}$ plane (lower panels) with $m_a$ fixed at $10^{-6}$ (left) and $0.1$ (right) MeV. The shaded regions indicate the sensitivities of CONUS and MINER experiments. The sensitivities of RECODE are shown by solid/dashed lines. Along with the expected 2 cpkkd background, two more conservative choices with 10 and 100 cpkkd are also used for the background. When the masses of ALP and dark photon are close, the decay length is also limited by the decay phase space, the sensitivity regions have sharp edge when the two masses are nearly degenerate as shown in the right panels. While when the mass difference is large, with larger couplings at mass slightly below MeV scale, the decay length is also too small for the ALP or dark photon to reach the detector. Hence there will be an upper and lower boundary for the exclusion region.

\begin{figure}[!tbp]
\centering
    \includegraphics[width=0.6\textwidth]{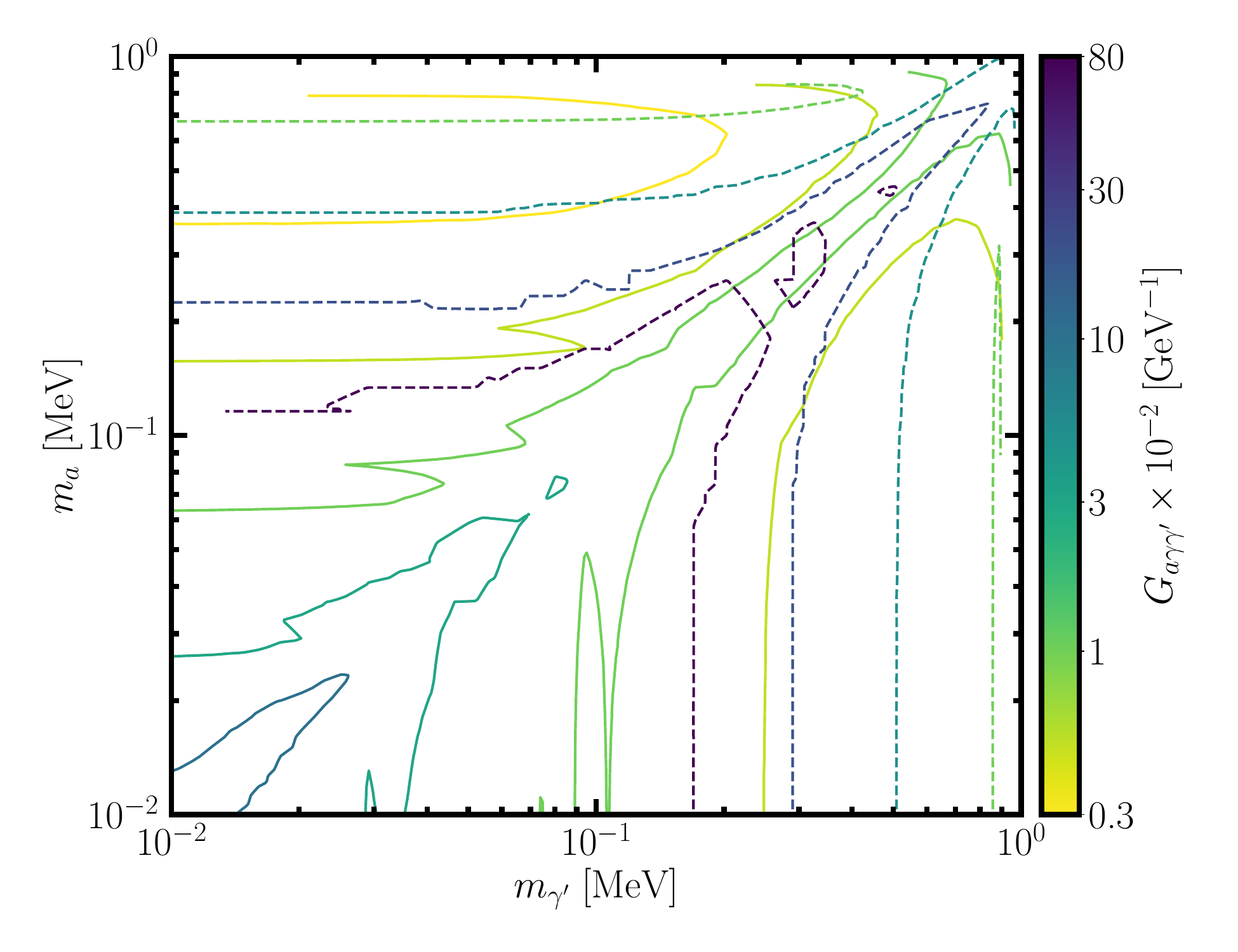}
\caption{The projected sensitivity at 95\% C.L. of the RECODE experiment with one year operation to axion and dark photon decay events. The solid curve denotes the lower boundary of the excluded coupling strength for the corresponding mass parameters, while the dashed curve indicates the upper boundary.}
\label{fig:Exclusion_Limit_Decay}
\end{figure}

The exclusion region at 95\% for $G_{a\gamma\gamma'}$ in $m_{\gamma'}$-$m_a$ from RECODE with 2 cpkkd background and one year operation is given in~\autoref{fig:Exclusion_Limit_Decay}, where the solid line indicates the lower boundary of the excluded region of the coupling for given masses, while the dashed line indicates the upper boundary (if smaller than $1\, {\rm GeV}^{-1}$) for the given masses. It is clear the exclusion sensitivity will be stronger at larger mass difference. When ALP and dark photon are degenerate, there will be no sensitivity from decay events.

\subsubsection{Detection through the Scattering}

The ALP and dark photon produced from the reactor can also be detected through their scattering with the nucleons ($aN\to\gamma'N$ and $\gamma'N\to aN$) within the detector. The event rate from scattering is given by
\begin{align}
    \frac{dN_\varphi^S}{dE_{nR}} = T \frac{m_{\rm det}N_T}{4\pi L^2}\int \frac{d\sigma}{dE_{nR}} \frac{dR_\varphi}{dE_\varphi}\exp\left(-\frac{L}{\beta\gamma c\tau}\right)dE_\varphi,
\end{align}
where $m_{\rm det}$ denotes the detector mass in $\rm kg$, $N_T = N_A/m_{\rm molar}$ with $N_A = 6.022\times10^{23}\,\rm mol^{-1}$ and $m_{\rm molar}$ the molar mass of the detector (in $\rm kg/mol$). $\frac{d\sigma}{dE_R}$ is the differential cross section for either $aN\to\gamma'N$ or $\gamma'N\to aN$ with $E_{nR}$ the nuclear recoil energy. For the germernium detector, we consider the Lindhard quenching factor (QF) converting the nuclear recoil energy to the energy detected:
\begin{align}
    Q(E) = \frac{\kappa g(\epsilon)}{1+\kappa g(\epsilon)},
\end{align}
with
\begin{align}
    \kappa &\approx 0.133 N_p^{2/3}(N_p+N_n)^{-1/2},\\
    g(\epsilon) &\approx 3\epsilon^{0.15}+0.7\epsilon^{0.6}+\epsilon, \\
    \epsilon &= 11.5 \left(\frac{E}{\rm keV}\right)N_p^{-7/3}.
\end{align}

\begin{figure}[!tbp]
\centering
\includegraphics[width=0.48\textwidth]{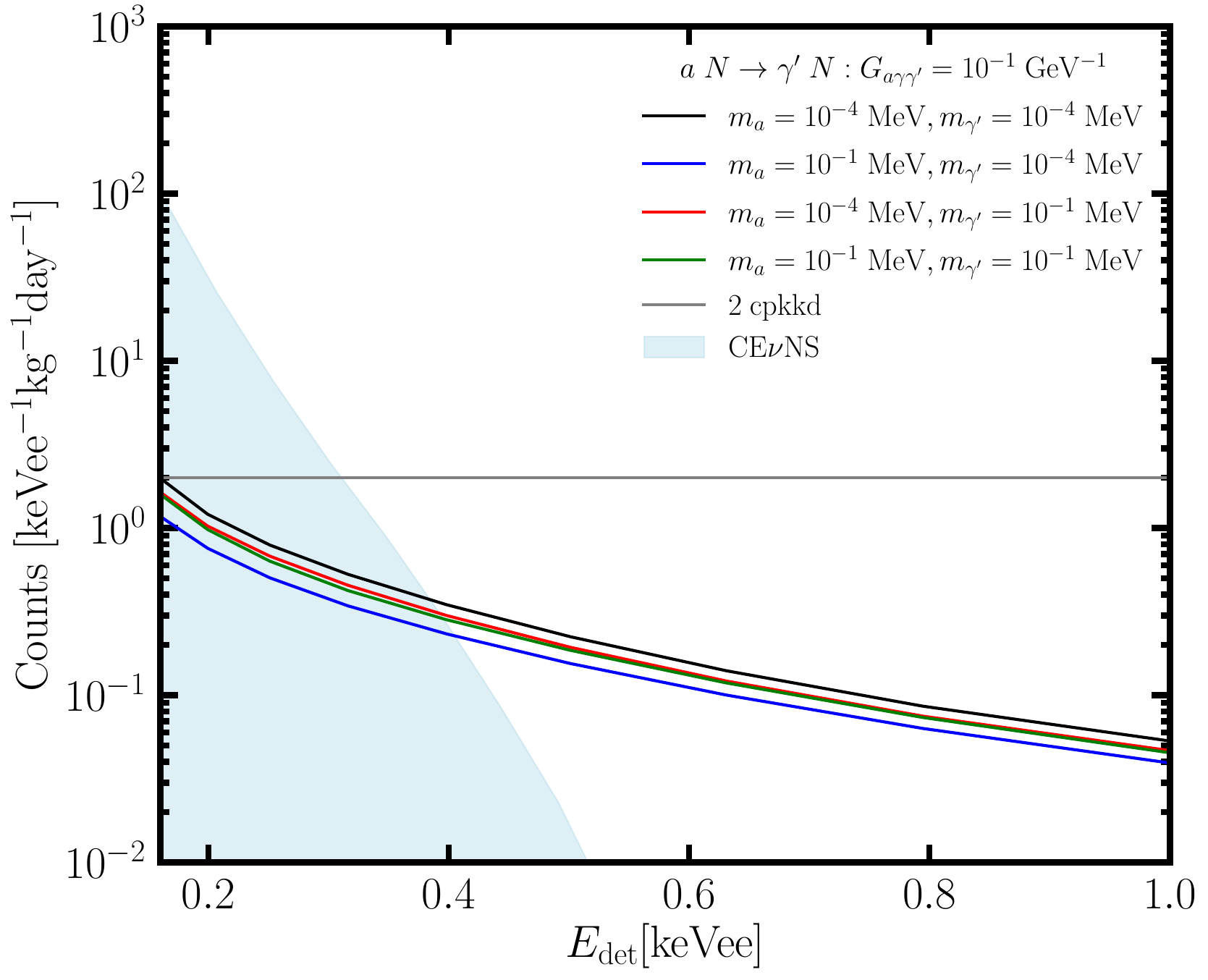}
\includegraphics[width=0.48\textwidth]{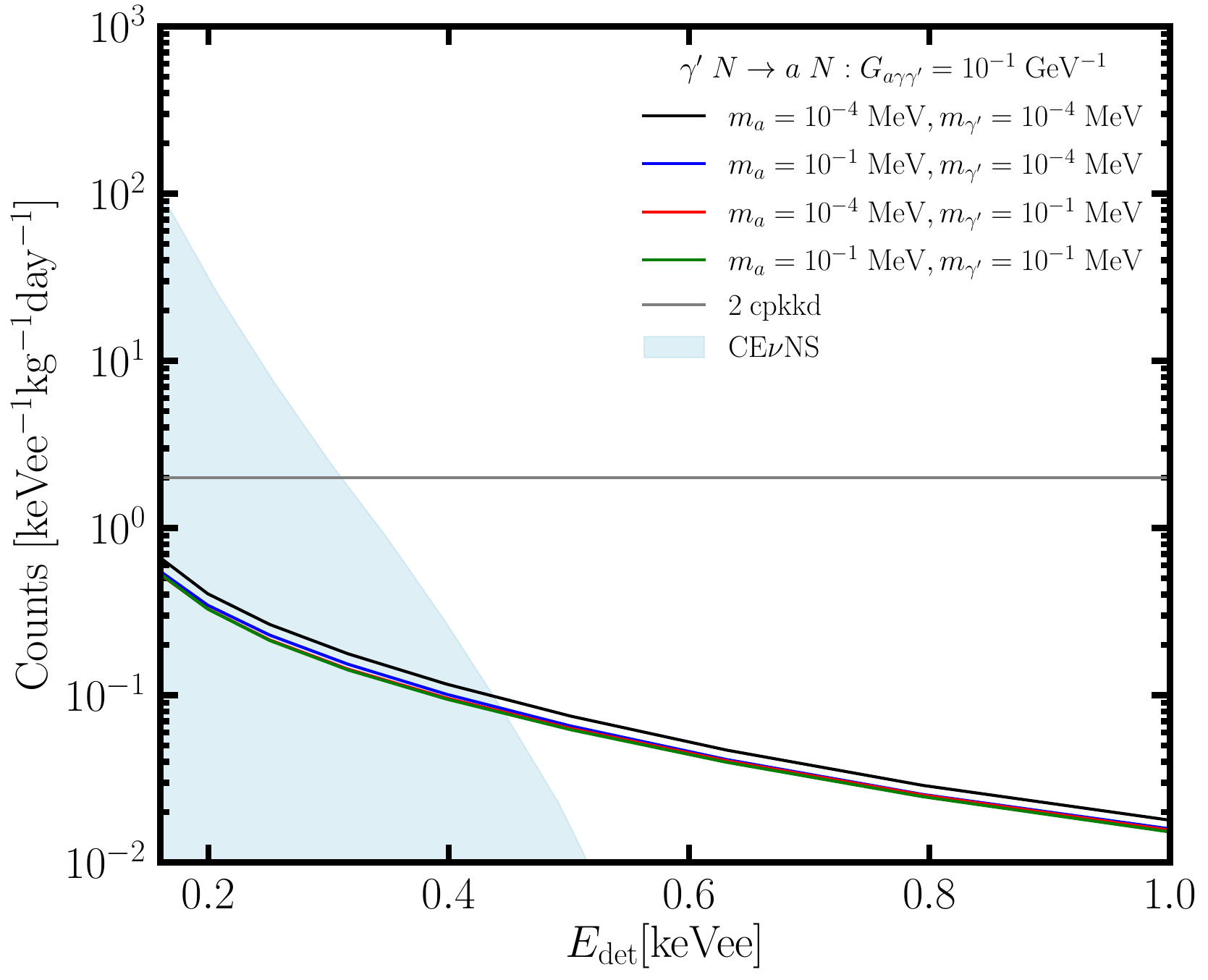}
\caption{
The recoil energy spectrum for $G_{a\gamma\gamma'}=10^{-1}\,{\rm GeV}^{-1}$ with five different mass parameter combinations for ALP scattering (left) and dark photon scattering (right). The blue shadow represents the reactor \cevns{} process contributions and the gray line indicates the expected 2 cpkkd background.
}
\label{fig:rate_spectrum_scattering}
\end{figure}

\begin{figure}[!tbp]
\centering
    \includegraphics[width=0.6\textwidth]{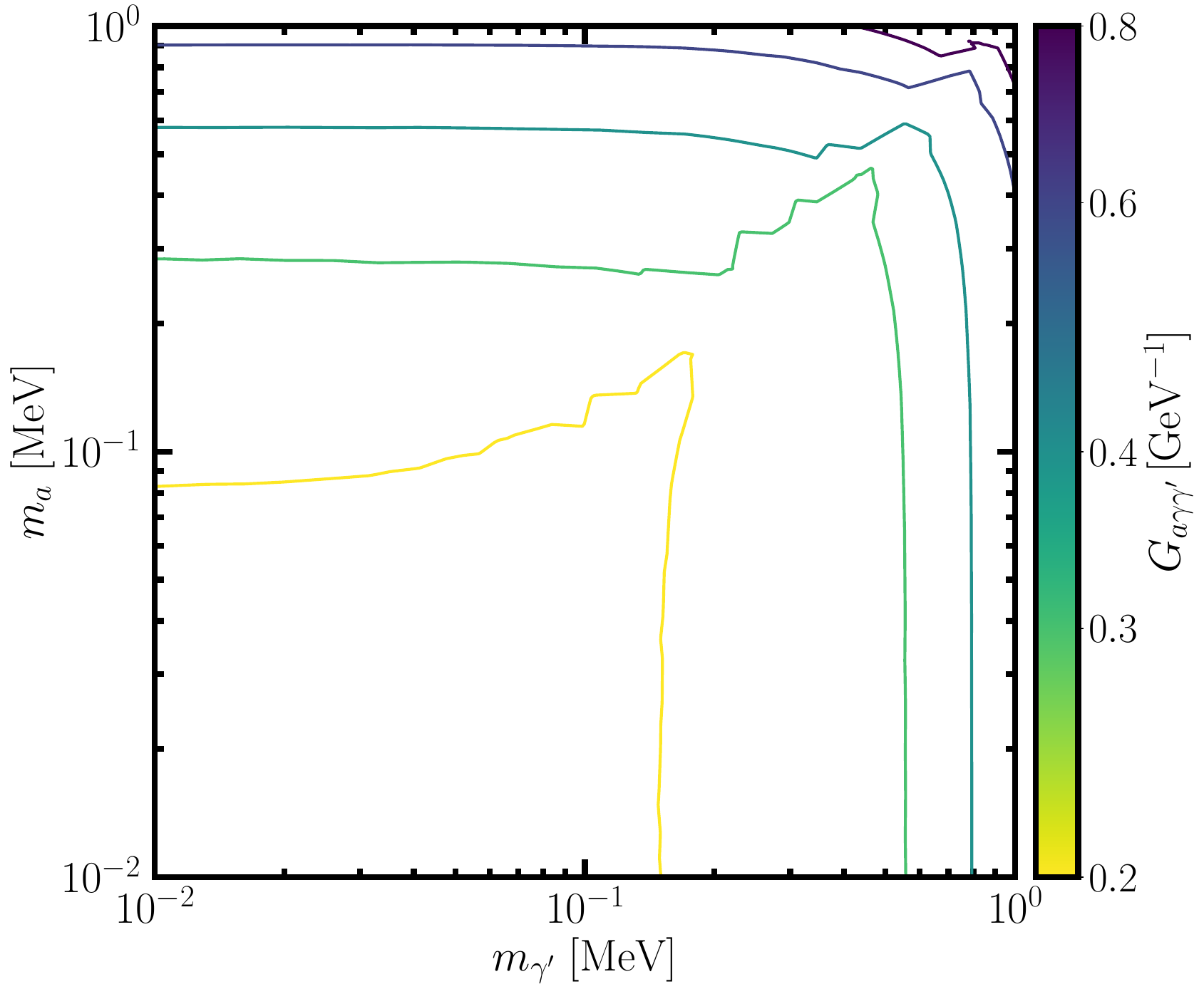}
    \caption{The upper bound on the $G_{a\gamma\gamma'}$ at 95\% C.L. of the RECODE experiment with one year operation from the scattering processes.}
\label{fig:Exclusion_Limit_scattering}
\end{figure}

The event spectra from the scattering for different choices of the masses with detector resolution taking into account are shown in~\autoref{fig:rate_spectrum_scattering}, where we also indicate the event spectrum from \cevns{} process induced by the neutrino from the reactor and the expected 2 cpkkd backgrounds. It is clear that the scattering spectrum is much lower than those induced from the ALP and dark photon decays under the same coupling $G_{a\gamma\gamma'}$. On the other hand, the induced energy from the scattering is also lower than that from the decay, and is within the similar region of the \cevns{} process. Hence, the detection of the event induced by the scattering process will be much harder. However, it is important in two kinds of regions: 1. the region with ALP and dark photon nearly degenerate where the heavier one will be long-lived; 2. the region with large coupling where the heavier one will be very shot-lived. In both regions, detection of decay events can be hard, the scattering process can thus be complementary to the decay process. The constraint on the coupling from scattering process only in the $m_{\gamma'}$-$m_a$ plane is shown in~\autoref{fig:Exclusion_Limit_scattering} where the solid lines indicate the 95\% C.L. upper bound on the coupling. It is clear that although the constraint is relatively weak compared with the results from decay events, it can be complemetary in the degenrate region.

\subsubsection{Combinations and Comparison with other Constraints}

\begin{figure}[!tbp]
\centering
    \includegraphics[width=0.6\textwidth]{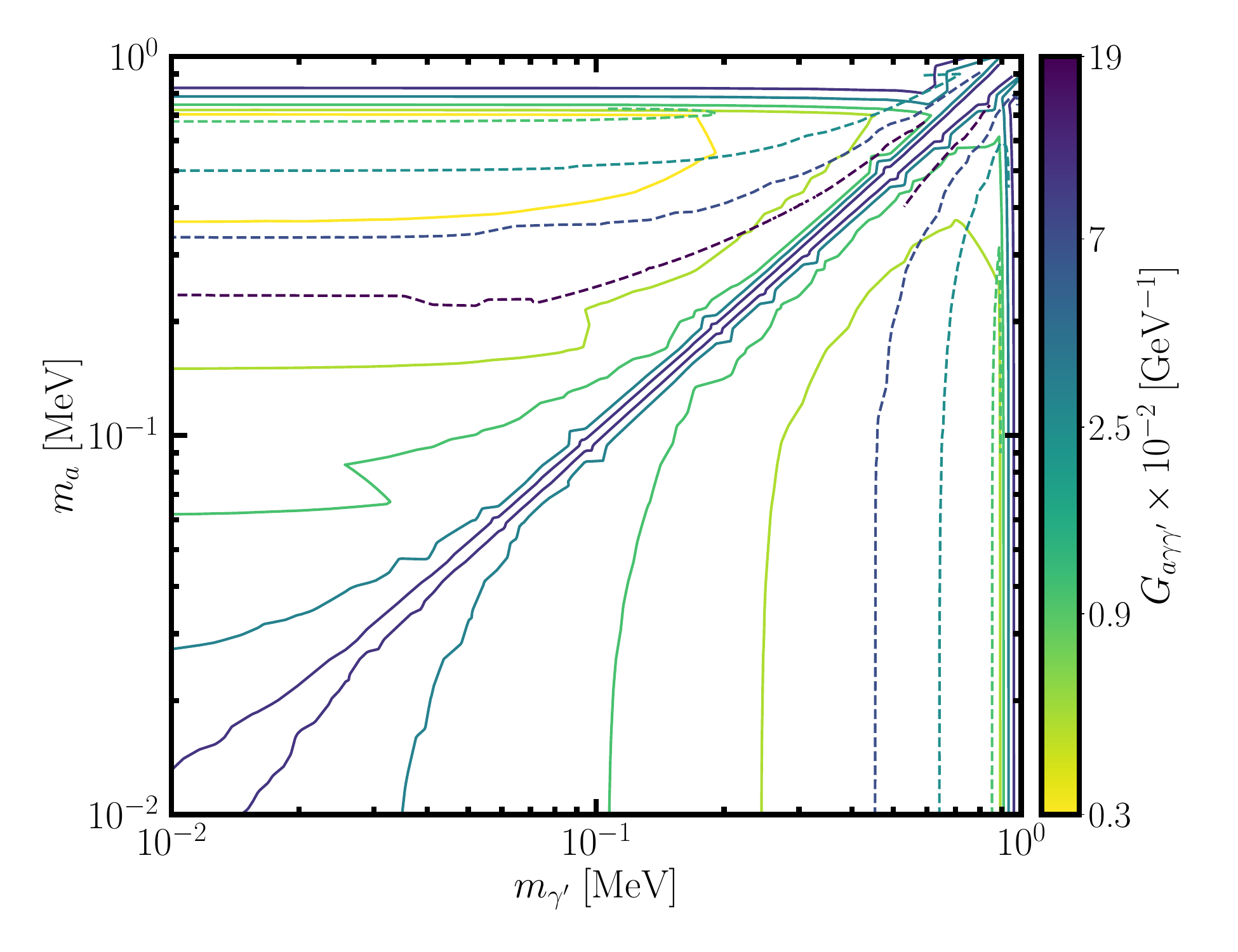}
\caption{The sensitivity at 95\% C.L. of the RECODE experiment with one year operation combining both decay and scattering events on $G_{a\gamma\gamma'}$ in $m_{\gamma'}$-$m_a$ plane. Solid lines denote the lower boundary of the excluded coupling strength for the corresponding mass parameters, while dashed lines denote the upper boundary.}
\label{fig:Exclusion_Limit_All}
\end{figure}

The combined excluded region is obtained by adding the contributions from both decay and scattering events. Note that the energy deposited in the detector is different for these two contributions, the total $\chi^2$ is a simple summation of the individual one from both decay and scattering analysis.
The excluded range of $G_{a\gamma\gamma'}$ is shown in~\autoref{fig:Exclusion_Limit_All} in the $m_{\gamma'}-m_a$ plane where the solid/dashed lines indicate the lower/upper boundary of the exclusion range of $G_{a\gamma\gamma'}$. The exclusion limit is mostly dominated by the decay processes, with the excpetion along the degenerate region ($m_{\gamma'}\approx m_a$) where the decay process does not provide sufficient event rate. In such regions, the scattering processes clearly provide a complementary contributions. Overall, the lower boundary of the excluded range of the coupling decreases with the increase of the mass before around 1 MeV, while the upper boundary decrease slightly more significantly with the increase of the mass. Further, the upper boundary will be larger than $\mathcal{O}(1\,\rm GeV^{-1})$ for mass smaller than about $\mathcal{O}(0.1)\,\rm MeV$ and hence is not considered in current analysis.

\begin{figure}[!tbp]
    \centering
    \includegraphics[width=0.48\textwidth]{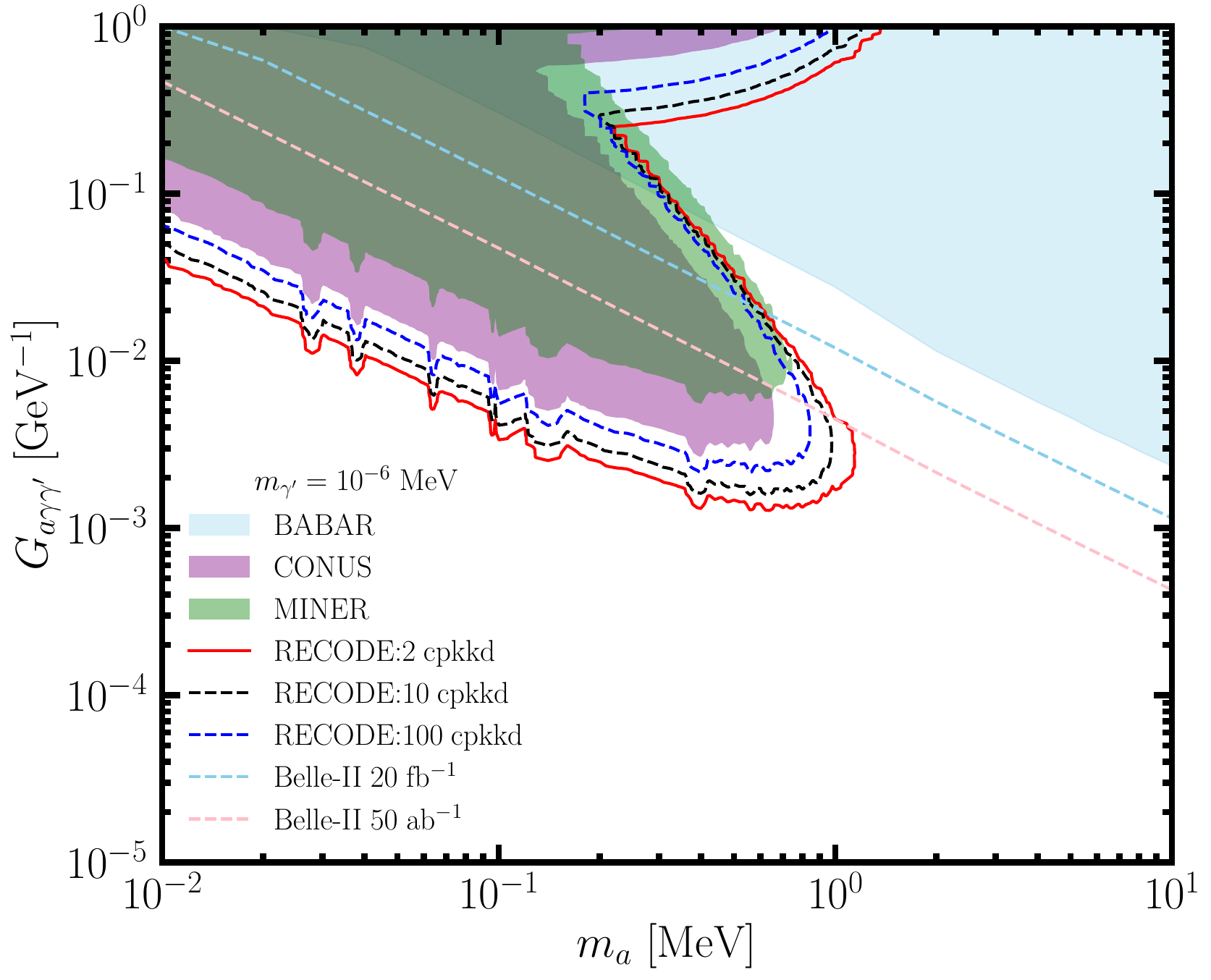}
    \includegraphics[width=0.48\textwidth]{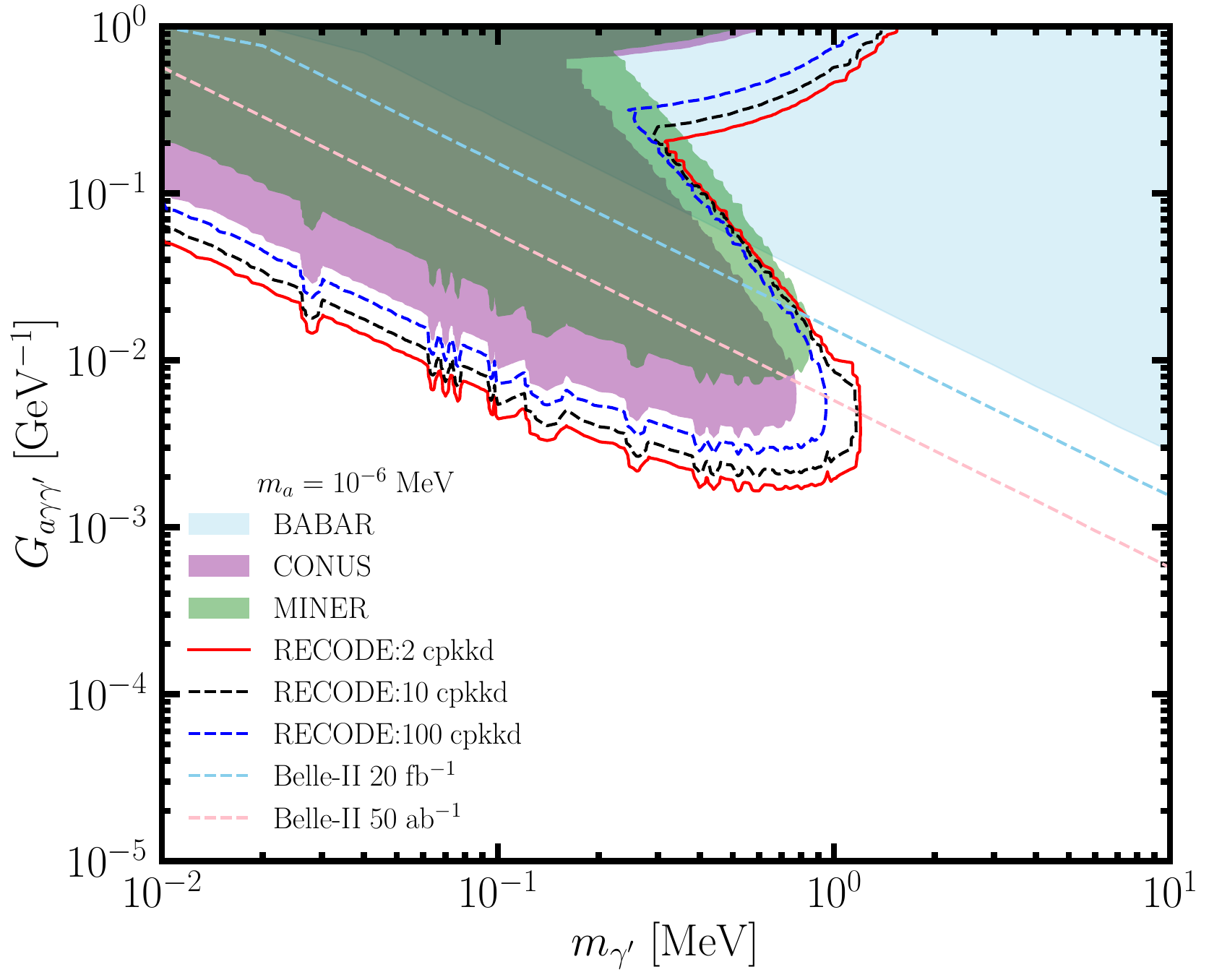}
    \caption{The sensitivity of \cevns{} experiments with one year operation compared with BABAR and Belle-II with either $m_{\gamma'}\ll m_a$ (left) and $m_a\ll m_{\gamma'}$ (right).}
    \label{fig:comparison}
\end{figure}

On the other hand, there are contraints from many other experiments including B factories, beam dump experiments and fixed target neutrino experiments. Here we focus on those constraints around MeV scale as discussed in~\cite{deNiverville:2018hrc,deNiverville:2019xsx,Essig:2013vha}, for a more comprehensive discussion on the dark axion portal with much smaller mass scale, we refer to Ref.~\cite{Arias:2020tzl}. To compared the sensitivity of B factories with the CEvNS experiments, we focus on two scenarios with either $m_{\gamma'}\ll m_a$ or $m_a\ll m_{\gamma'}$ as shown in~\autoref{fig:comparison} where the exclusion regions from CEvNS experiments with one year operation include the contributions from both decay and scattering events. Compared with the results sorely from decay events as shown in~\autoref{fig:Ndecay_Ga_ma_dark_axion_portal}, it is clear that the scattering contributions can also be complementary in the regions with large masses and couplings. On the other hand, we also recast the sensitivity of the BaBar experiment~\cite{BaBar:2001yhh,BaBar:2008aby} indicated by the light blue region and estimate the prospects of the Belle-II experiment~\cite{Belle-II:2010dht} indicated by the light blue/orange dashed lines.
We find that although collider experiments exhibit excellent exclusion power in the GeV mass range, their sensitivity rapidly deteriorates as the masses of the axion and the dark photon decrease. In this regime, the lifetimes of these particles become significantly longer, making them more likely to escape detectors with characteristic sizes of $\mathcal{O}(1\,\mathrm{m})$. Although reactor experiments also employ detectors of relatively modest size, their typical energy scale lies in the sub-MeV region. Consequently, reactor experiments can provide stronger constraints on axions and dark photons in this mass range compared with collider experiments.

\section{Summary}
\label{sec:Conclusion}

Currently, many experimental studies utilize nuclear reactors to search for weakly interacting particles such as neutrinos, axions (or axion-like particles, ALPs), and dark photons. Nuclear reactors serve as intense sources of MeV scale neutrinos as well as photons, making them well-suited for ALP production. The high flux of these particles not only enhances the precision of neutrino measurements but also improves the sensitivity to rare processes involving feebly interacting particles. This makes reactor-based experiments particularly advantageous for probing new physics, especially in the search for dark sector particles in the MeV mass range. In this context, the dark axion portal, which facilitates interactions between SM photons, dark photons, and axions, provides a compelling framework for investigation within the \cevns{} experiments at nuclear reactors.

In this study, we investigate the potential of the reactor \cevns{} experiments, in particular the RECODE experiment, to probe the axion and dark photon from dark axion portal. We considered both decay and scattering events for the detection. Although, in most parameter space, the sensitivity is dominated by the decay events, the scattering processes can be complementary when the masses of ALP and dark photon are degenerate and also in the region where one of the masses and the coupling are large. Combining both decay and scattering channels, we can excluded as low as $\mathcal{O}(10^{-3})$ for the dark axion portal coupling $G_{a\gamma\gamma'}$ with one year operation. With more data accumulated, the sensivity will increase accordingly. Compared with the constraints from B factories which dominate at higher mass region, it is shown that the reactor \cevns{} experiments can be complementary in the sub-MeV region demonstrating the unique role of reactor \cevns{} experiments for probing the new physics around MeV scale.

\begin{acknowledgments}
This work is supported by the National Key Research and Development Program of China (Grant No. 2022YFA1605000). YJS and YW are also supported by the National Natural Science Foundation of China (NNSFC) under grant No.~12305112. LTY is supported by the NNSFC under grant No. 12322511. JT is supported by the NNSFC under grant No. 12175038. The authors gratefully acknowledge the valuable discussions and insights provided by the members of the China Collaboration of Precision Testing and New Physics (CPTNP).
\end{acknowledgments}

\bibliographystyle{JHEP}
\bibliography{references}

\providecommand{\href}[2]{#2}\begingroup\raggedright\begin{thebibliography}{10}

\bibitem{Peccei:1977hh}
R.D.~Peccei and H.R.~Quinn, \emph{{CP Conservation in the Presence of Instantons}}, \href{https://doi.org/10.1103/PhysRevLett.38.1440}{\emph{Phys. Rev. Lett.} {\bfseries 38} (1977) 1440}.

\bibitem{Wilczek:1977pj}
F.~Wilczek, \emph{{Problem of Strong $P$ and $T$ Invariance in the Presence of Instantons}}, \href{https://doi.org/10.1103/PhysRevLett.40.279}{\emph{Phys. Rev. Lett.} {\bfseries 40} (1978) 279}.

\bibitem{Weinberg:1977ma}
S.~Weinberg, \emph{{A New Light Boson?}}, \href{https://doi.org/10.1103/PhysRevLett.40.223}{\emph{Phys. Rev. Lett.} {\bfseries 40} (1978) 223}.

\bibitem{Peccei:1977ur}
R.D.~Peccei and H.R.~Quinn, \emph{{Constraints Imposed by CP Conservation in the Presence of Instantons}}, \href{https://doi.org/10.1103/PhysRevD.16.1791}{\emph{Phys. Rev. D} {\bfseries 16} (1977) 1791}.

\bibitem{Duffy:2009ig}
L.D.~Duffy and K.~van Bibber, \emph{{Axions as Dark Matter Particles}}, \href{https://doi.org/10.1088/1367-2630/11/10/105008}{\emph{New J. Phys.} {\bfseries 11} (2009) 105008} [\href{https://arxiv.org/abs/0904.3346}{{\ttfamily 0904.3346}}].

\bibitem{Marsh_2016}
D.J.~Marsh, \emph{Axion cosmology}, \href{https://doi.org/10.1016/j.physrep.2016.06.005}{\emph{Physics Reports} {\bfseries 643} (2016) 1–79}.

\bibitem{Battaglieri:2017aum}
M.~Battaglieri et~al., \emph{{US Cosmic Visions: New Ideas in Dark Matter 2017: Community Report}},  in \emph{{U.S. Cosmic Visions: New Ideas in Dark Matter}}, 7, 2017 [\href{https://arxiv.org/abs/1707.04591}{{\ttfamily 1707.04591}}].

\bibitem{Donnelly:1978ty}
T.W.~Donnelly, S.J.~Freedman, R.S.~Lytel, R.D.~Peccei and M.~Schwartz, \emph{{Do Axions Exist?}}, \href{https://doi.org/10.1103/PhysRevD.18.1607}{\emph{Phys. Rev. D} {\bfseries 18} (1978) 1607}.

\bibitem{Vuilleumier:1981dq}
J.L.~Vuilleumier, F.~Boehm, A.A.~Hahn, H.~Kwon, F.~Von~Feilitzsch and R.L.~Mossbauer, \emph{{An Experimental Limit on Production of Axions in a Fission Reactor}}, \href{https://doi.org/10.1016/0370-2693(81)90059-9}{\emph{Phys. Lett. B} {\bfseries 101} (1981) 341}.

\bibitem{Datar:1982ef}
V.M.~Datar, C.V.K.~Baba, M.G.~Betigeri and P.~Singh, \emph{{SEARCH FOR AXIONS IN THERMAL NEUTRON CAPTURE BY PROTONS}}, \href{https://doi.org/10.1016/0370-2693(82)90016-8}{\emph{Phys. Lett. B} {\bfseries 114} (1982) 63}.

\bibitem{Koch:1986aq}
H.R.~Koch and O.W.B.~Schult, \emph{{SEARCH FOR AXIONS AT THE NUCLEAR POWER REACTOR BIBLIS-A}}, \href{https://doi.org/10.1007/BF02776999}{\emph{Nuovo Cim. A} {\bfseries 96} (1986) 182}.

\bibitem{Altmann:1995bw}
M.~Altmann, F.~von Feilitzsch, C.~Hagner, L.~Oberauer, Y.~Declais and E.~Kajfasz, \emph{{Search for the electron positron decay of axions and axion - like particles at a nuclear power reactor at Bugey}}, \href{https://doi.org/10.1007/BF01566670}{\emph{Z. Phys. C} {\bfseries 68} (1995) 221}.

\bibitem{Kim:1979if}
J.E.~Kim, \emph{{Weak Interaction Singlet and Strong CP Invariance}}, \href{https://doi.org/10.1103/PhysRevLett.43.103}{\emph{Phys. Rev. Lett.} {\bfseries 43} (1979) 103}.

\bibitem{Shifman:1979if}
M.A.~Shifman, A.I.~Vainshtein and V.I.~Zakharov, \emph{{Can Confinement Ensure Natural CP Invariance of Strong Interactions?}}, \href{https://doi.org/10.1016/0550-3213(80)90209-6}{\emph{Nucl. Phys. B} {\bfseries 166} (1980) 493}.

\bibitem{Dine:1981rt}
M.~Dine, W.~Fischler and M.~Srednicki, \emph{{A Simple Solution to the Strong CP Problem with a Harmless Axion}}, \href{https://doi.org/10.1016/0370-2693(81)90590-6}{\emph{Phys. Lett. B} {\bfseries 104} (1981) 199}.

\bibitem{TEXONO:2006spf}
{\scshape TEXONO} collaboration, \emph{{Search of axions at the Kuo-Sheng nuclear power station with a high-purity germanium detector}}, \href{https://doi.org/10.1103/PhysRevD.75.052004}{\emph{Phys. Rev. D} {\bfseries 75} (2007) 052004} [\href{https://arxiv.org/abs/hep-ex/0609001}{{\ttfamily hep-ex/0609001}}].

\bibitem{Holdom:1985ag}
B.~Holdom, \emph{{Two U(1)'s and Epsilon Charge Shifts}}, \href{https://doi.org/10.1016/0370-2693(86)91377-8}{\emph{Phys. Lett. B} {\bfseries 166} (1986) 196}.

\bibitem{Arkani-Hamed:2008kxc}
N.~Arkani-Hamed and N.~Weiner, \emph{{LHC Signals for a SuperUnified Theory of Dark Matter}}, \href{https://doi.org/10.1088/1126-6708/2008/12/104}{\emph{JHEP} {\bfseries 12} (2008) 104} [\href{https://arxiv.org/abs/0810.0714}{{\ttfamily 0810.0714}}].

\bibitem{PhysRevD.80.015003}
R.~Essig, P.~Schuster and N.~Toro, \emph{Probing dark forces and light hidden sectors at low-energy ${e}^{+}{e}^{\ensuremath{-}}$ colliders}, \href{https://doi.org/10.1103/PhysRevD.80.015003}{\emph{Phys. Rev. D} {\bfseries 80} (2009) 015003}.

\bibitem{Goodsell:2011wn}
M.~Goodsell, S.~Ramos-Sanchez and A.~Ringwald, \emph{{Kinetic Mixing of U(1)s in Heterotic Orbifolds}}, \href{https://doi.org/10.1007/JHEP01(2012)021}{\emph{JHEP} {\bfseries 01} (2012) 021} [\href{https://arxiv.org/abs/1110.6901}{{\ttfamily 1110.6901}}].

\bibitem{Alexander:2016aln}
J.~Alexander et~al., \emph{{Dark Sectors 2016 Workshop: Community Report}},  8, 2016 [\href{https://arxiv.org/abs/1608.08632}{{\ttfamily 1608.08632}}].

\bibitem{Essig:2013lka}
R.~Essig et~al., \emph{{Working Group Report: New Light Weakly Coupled Particles}},  in \emph{{Snowmass 2013}: {Snowmass on the Mississippi}}, 10, 2013 [\href{https://arxiv.org/abs/1311.0029}{{\ttfamily 1311.0029}}].

\bibitem{Kaneta:2016wvf}
K.~Kaneta, H.-S.~Lee and S.~Yun, \emph{{Portal Connecting Dark Photons and Axions}}, \href{https://doi.org/10.1103/PhysRevLett.118.101802}{\emph{Phys. Rev. Lett.} {\bfseries 118} (2017) 101802} [\href{https://arxiv.org/abs/1611.01466}{{\ttfamily 1611.01466}}].

\bibitem{Kaneta:2017wfh}
K.~Kaneta, H.-S.~Lee and S.~Yun, \emph{{Dark photon relic dark matter production through the dark axion portal}}, \href{https://doi.org/10.1103/PhysRevD.95.115032}{\emph{Phys. Rev. D} {\bfseries 95} (2017) 115032} [\href{https://arxiv.org/abs/1704.07542}{{\ttfamily 1704.07542}}].

\bibitem{Choi:2018dqr}
K.~Choi, H.~Kim and T.~Sekiguchi, \emph{{Late-Time Magnetogenesis Driven by Axionlike Particle Dark Matter and a Dark Photon}}, \href{https://doi.org/10.1103/PhysRevLett.121.031102}{\emph{Phys. Rev. Lett.} {\bfseries 121} (2018) 031102} [\href{https://arxiv.org/abs/1802.07269}{{\ttfamily 1802.07269}}].

\bibitem{deNiverville:2018hrc}
P.~deNiverville, H.-S.~Lee and M.-S.~Seo, \emph{{Implications of the dark axion portal for the muon g{\ensuremath{-}}2 , B factories, fixed target neutrino experiments, and beam dumps}}, \href{https://doi.org/10.1103/PhysRevD.98.115011}{\emph{Phys. Rev. D} {\bfseries 98} (2018) 115011} [\href{https://arxiv.org/abs/1806.00757}{{\ttfamily 1806.00757}}].

\bibitem{BaBar:2001yhh}
{\scshape BaBar} collaboration, \emph{{The BaBar detector}}, \href{https://doi.org/10.1016/S0168-9002(01)02012-5}{\emph{Nucl. Instrum. Meth. A} {\bfseries 479} (2002) 1} [\href{https://arxiv.org/abs/hep-ex/0105044}{{\ttfamily hep-ex/0105044}}].

\bibitem{Belle-II:2010dht}
{\scshape Belle-II} collaboration, \emph{{Belle II Technical Design Report}},  \href{https://arxiv.org/abs/1011.0352}{{\ttfamily 1011.0352}}.

\bibitem{NEOS:2016wee}
{\scshape NEOS} collaboration, \emph{{Sterile Neutrino Search at the NEOS Experiment}}, \href{https://doi.org/10.1103/PhysRevLett.118.121802}{\emph{Phys. Rev. Lett.} {\bfseries 118} (2017) 121802} [\href{https://arxiv.org/abs/1610.05134}{{\ttfamily 1610.05134}}].

\bibitem{Proceedings:2019mnq}
\emph{{Proceedings of The Magnificent CE$\nu$NS Workshop 2018}}, 10, 2019.
\newblock 10.5281/zenodo.3489190.

\bibitem{RENO:2010vlj}
{\scshape RENO} collaboration, \emph{{RENO: An Experiment for Neutrino Oscillation Parameter $\theta_{13}$ Using Reactor Neutrinos at Yonggwang}},  \href{https://arxiv.org/abs/1003.1391}{{\ttfamily 1003.1391}}.

\bibitem{LSND:1996jxj}
{\scshape LSND} collaboration, \emph{{The Liquid scintillator neutrino detector and LAMPF neutrino source}}, \href{https://doi.org/10.1016/S0168-9002(96)01155-2}{\emph{Nucl. Instrum. Meth. A} {\bfseries 388} (1997) 149} [\href{https://arxiv.org/abs/nucl-ex/9605002}{{\ttfamily nucl-ex/9605002}}].

\bibitem{MiniBooNE:2013dds}
{\scshape MiniBooNE} collaboration, \emph{{Measurement of the Antineutrino Neutral-Current Elastic Differential Cross Section}}, \href{https://doi.org/10.1103/PhysRevD.91.012004}{\emph{Phys. Rev. D} {\bfseries 91} (2015) 012004} [\href{https://arxiv.org/abs/1309.7257}{{\ttfamily 1309.7257}}].

\bibitem{Gninenko:2011uv}
S.N.~Gninenko, \emph{{Stringent limits on the $\pi^0 \to \gamma X, X \to e+e-$ decay from neutrino experiments and constraints on new light gauge bosons}}, \href{https://doi.org/10.1103/PhysRevD.85.055027}{\emph{Phys. Rev. D} {\bfseries 85} (2012) 055027} [\href{https://arxiv.org/abs/1112.5438}{{\ttfamily 1112.5438}}].

\bibitem{Freedman:1973yd}
D.Z.~Freedman, \emph{{Coherent Neutrino Nucleus Scattering as a Probe of the Weak Neutral Current}}, \href{https://doi.org/10.1103/PhysRevD.9.1389}{\emph{Phys. Rev. D} {\bfseries 9} (1974) 1389}.

\bibitem{Abdullah:2022zue}
M.~Abdullah et~al., \emph{{Coherent elastic neutrino-nucleus scattering: Terrestrial and astrophysical applications}},  \href{https://arxiv.org/abs/2203.07361}{{\ttfamily 2203.07361}}.

\bibitem{AristizabalSierra:2019ykk}
D.~Aristizabal~Sierra, B.~Dutta, S.~Liao and L.E.~Strigari, \emph{{Coherent elastic neutrino-nucleus scattering in multi-ton scale dark matter experiments: Classification of vector and scalar interactions new physics signals}}, \href{https://doi.org/10.1007/JHEP12(2019)124}{\emph{JHEP} {\bfseries 12} (2019) 124} [\href{https://arxiv.org/abs/1910.12437}{{\ttfamily 1910.12437}}].

\bibitem{ParticleDataGroup:2018ovx}
{\scshape Particle Data Group} collaboration, \emph{{Review of Particle Physics}}, \href{https://doi.org/10.1103/PhysRevD.98.030001}{\emph{Phys. Rev. D} {\bfseries 98} (2018) 030001}.

\bibitem{Safronova:2017xyt}
M.S.~Safronova, D.~Budker, D.~DeMille, D.F.J.~Kimball, A.~Derevianko and C.W.~Clark, \emph{{Search for New Physics with Atoms and Molecules}}, \href{https://doi.org/10.1103/RevModPhys.90.025008}{\emph{Rev. Mod. Phys.} {\bfseries 90} (2018) 025008} [\href{https://arxiv.org/abs/1710.01833}{{\ttfamily 1710.01833}}].

\bibitem{Cadeddu:2021dqx}
M.~Cadeddu, N.~Cargioli, F.~Dordei, C.~Giunti and E.~Picciau, \emph{{Muon and electron g-2 and proton and cesium weak charges implications on dark Zd models}}, \href{https://doi.org/10.1103/PhysRevD.104.L011701}{\emph{Phys. Rev. D} {\bfseries 104} (2021) 011701} [\href{https://arxiv.org/abs/2104.03280}{{\ttfamily 2104.03280}}].

\bibitem{Majumdar:2022nby}
A.~Majumdar, D.K.~Papoulias, R.~Srivastava and J.W.F.~Valle, \emph{{Physics implications of recent Dresden-II reactor data}}, \href{https://doi.org/10.1103/PhysRevD.106.093010}{\emph{Phys. Rev. D} {\bfseries 106} (2022) 093010} [\href{https://arxiv.org/abs/2208.13262}{{\ttfamily 2208.13262}}].

\bibitem{XENON:2024hup}
{\scshape XENON} collaboration, \emph{{First Search for Light Dark Matter in the Neutrino Fog with XENONnT}}, \href{https://doi.org/10.1103/PhysRevLett.134.111802}{\emph{Phys. Rev. Lett.} {\bfseries 134} (2025) 111802} [\href{https://arxiv.org/abs/2409.17868}{{\ttfamily 2409.17868}}].

\bibitem{DeRomeri:2025nkx}
V.~De~Romeri, A.~Majumdar, D.K.~Papoulias and R.~Srivastava, \emph{{New light mediators and the neutrino fog: Implications from XENONnT nuclear recoil data}},  \href{https://arxiv.org/abs/2512.08853}{{\ttfamily 2512.08853}}.

\bibitem{Billard:2013qya}
J.~Billard, L.~Strigari and E.~Figueroa-Feliciano, \emph{{Implication of neutrino backgrounds on the reach of next generation dark matter direct detection experiments}}, \href{https://doi.org/10.1103/PhysRevD.89.023524}{\emph{Phys. Rev. D} {\bfseries 89} (2014) 023524} [\href{https://arxiv.org/abs/1307.5458}{{\ttfamily 1307.5458}}].

\bibitem{COHERENT:2017ipa}
{\scshape COHERENT} collaboration, \emph{{Observation of Coherent Elastic Neutrino-Nucleus Scattering}}, \href{https://doi.org/10.1126/science.aao0990}{\emph{Science} {\bfseries 357} (2017) 1123} [\href{https://arxiv.org/abs/1708.01294}{{\ttfamily 1708.01294}}].

\bibitem{COHERENT:2021xmm}
{\scshape COHERENT} collaboration, \emph{{Measurement of the Coherent Elastic Neutrino-Nucleus Scattering Cross Section on CsI by COHERENT}}, \href{https://doi.org/10.1103/PhysRevLett.129.081801}{\emph{Phys. Rev. Lett.} {\bfseries 129} (2022) 081801} [\href{https://arxiv.org/abs/2110.07730}{{\ttfamily 2110.07730}}].

\bibitem{COHERENT:2020iec}
{\scshape COHERENT} collaboration, \emph{{First Measurement of Coherent Elastic Neutrino-Nucleus Scattering on Argon}}, \href{https://doi.org/10.1103/PhysRevLett.126.012002}{\emph{Phys. Rev. Lett.} {\bfseries 126} (2021) 012002} [\href{https://arxiv.org/abs/2003.10630}{{\ttfamily 2003.10630}}].

\bibitem{COHERENT:2026ewu}
{\scshape COHERENT} collaboration, \emph{{The COHERENT Experiment: 2026 Update}},  \href{https://arxiv.org/abs/2602.15652}{{\ttfamily 2602.15652}}.

\bibitem{XENON:2022ltv}
{\scshape XENON} collaboration, \emph{{Search for New Physics in Electronic Recoil Data from XENONnT}}, \href{https://doi.org/10.1103/PhysRevLett.129.161805}{\emph{Phys. Rev. Lett.} {\bfseries 129} (2022) 161805} [\href{https://arxiv.org/abs/2207.11330}{{\ttfamily 2207.11330}}].

\bibitem{XENON:2024ijk}
{\scshape XENON} collaboration, \emph{{First Indication of Solar B8 Neutrinos via Coherent Elastic Neutrino-Nucleus Scattering with XENONnT}}, \href{https://doi.org/10.1103/PhysRevLett.133.191002}{\emph{Phys. Rev. Lett.} {\bfseries 133} (2024) 191002} [\href{https://arxiv.org/abs/2408.02877}{{\ttfamily 2408.02877}}].

\bibitem{PandaX:2024muv}
{\scshape PandaX} collaboration, \emph{{First Indication of Solar B8 Neutrinos through Coherent Elastic Neutrino-Nucleus Scattering in PandaX-4T}}, \href{https://doi.org/10.1103/PhysRevLett.133.191001}{\emph{Phys. Rev. Lett.} {\bfseries 133} (2024) 191001} [\href{https://arxiv.org/abs/2407.10892}{{\ttfamily 2407.10892}}].

\bibitem{PandaX:2024cic}
{\scshape PandaX} collaboration, \emph{{Exploring New Physics with PandaX-4T Low Energy Electronic Recoil Data}}, \href{https://doi.org/10.1103/PhysRevLett.134.041001}{\emph{Phys. Rev. Lett.} {\bfseries 134} (2025) 041001} [\href{https://arxiv.org/abs/2408.07641}{{\ttfamily 2408.07641}}].

\bibitem{CONUS:2024lnu}
{\scshape CONUS+} collaboration, \emph{{CONUS+~Experiment}}, \href{https://doi.org/10.1140/epjc/s10052-024-13551-6}{\emph{Eur. Phys. J. C} {\bfseries 84} (2024) 1265} [\href{https://arxiv.org/abs/2407.11912}{{\ttfamily 2407.11912}}].

\bibitem{Colaresi:2022obx}
J.~Colaresi, J.I.~Collar, T.W.~Hossbach, C.M.~Lewis and K.M.~Yocum, \emph{{Measurement of Coherent Elastic Neutrino-Nucleus Scattering from Reactor Antineutrinos}}, \href{https://doi.org/10.1103/PhysRevLett.129.211802}{\emph{Phys. Rev. Lett.} {\bfseries 129} (2022) 211802} [\href{https://arxiv.org/abs/2202.09672}{{\ttfamily 2202.09672}}].

\bibitem{Akimov:2017hee}
D.Y.~Akimov et~al., \emph{{Status of the RED-100 experiment}}, \href{https://doi.org/10.1088/1748-0221/12/06/C06018}{\emph{JINST} {\bfseries 12} (2017) C06018}.

\bibitem{Strauss2017}
R.S.~et~al., \emph{The european physical journal c}, \href{https://doi.org/10.1140/epjc/s10052-017-5068-2}{\emph{Eur. Phys. J. C} {\bfseries 77} (2017) }.

\bibitem{MINER:2016igy}
{\scshape MINER} collaboration, \emph{{Background Studies for the MINER Coherent Neutrino Scattering Reactor Experiment}}, \href{https://doi.org/10.1016/j.nima.2017.02.024}{\emph{Nucl. Instrum. Meth. A} {\bfseries 853} (2017) 53} [\href{https://arxiv.org/abs/1609.02066}{{\ttfamily 1609.02066}}].

\bibitem{James_2019}
{James B. Dent, Bhaskar Dutta, Doojin Kim, Shu Liao, Rupak Mahapatra, Kuver Sinha, Adrian Thompson}, \emph{{New Directions for Axion Searches via Scattering at Reactor Neutrino Experiments}}, \href{https://doi.org/10.1103/PhysRevLett.124.211804}{\emph{Phys. Rev. Lett.} {\bfseries 124} (2020) 211804} [\href{https://arxiv.org/abs/1912.05733v2}{{\ttfamily 1912.05733v2}}].

\bibitem{CONNIE:2016ggr}
{\scshape CONNIE} collaboration, \emph{{The CONNIE experiment}}, \href{https://doi.org/10.1088/1742-6596/761/1/012057}{\emph{J. Phys. Conf. Ser.} {\bfseries 761} (2016) 012057} [\href{https://arxiv.org/abs/1608.01565}{{\ttfamily 1608.01565}}].

\bibitem{RELICS:2024opj}
{\scshape RELICS} collaboration, \emph{{Reactor neutrino liquid xenon coherent elastic scattering experiment}}, \href{https://doi.org/10.1103/PhysRevD.110.072011}{\emph{Phys. Rev. D} {\bfseries 110} (2024) 072011} [\href{https://arxiv.org/abs/2405.05554}{{\ttfamily 2405.05554}}].

\bibitem{Yang:2024exl}
L.T.~Yang, Y.F.~Liang and Q.~Yue, \emph{{RECODE program for reactor neutrino CEvNS detection with PPC Germanium detector}}, \href{https://doi.org/10.22323/1.441.0296}{\emph{PoS} {\bfseries TAUP2023} (2024) 296}.

\bibitem{Billard:2016giu}
J.~Billard et~al., \emph{{Coherent Neutrino Scattering with Low Temperature Bolometers at Chooz Reactor Complex}}, \href{https://doi.org/10.1088/1361-6471/aa83d0}{\emph{J. Phys. G} {\bfseries 44} (2017) 105101} [\href{https://arxiv.org/abs/1612.09035}{{\ttfamily 1612.09035}}].

\bibitem{Buck:2020opf}
C.~Buck et~al., \emph{{A novel experiment for coherent elastic neutrino nucleus scattering: CONUS}}, \href{https://doi.org/10.1088/1742-6596/1342/1/012094}{\emph{J. Phys. Conf. Ser.} {\bfseries 1342} (2020) 012094}.

\bibitem{Wong:2015kgl}
H.T.-K.~Wong, \emph{{Taiwan EXperiment On NeutrinO \textemdash{} History and Prospects}}, \href{https://doi.org/10.1142/S0217751X18300144}{\emph{The Universe} {\bfseries 3} (2015) 22} [\href{https://arxiv.org/abs/1608.00306}{{\ttfamily 1608.00306}}].

\bibitem{DeRomeri:2022twg}
V.~De~Romeri, O.G.~Miranda, D.K.~Papoulias, G.~Sanchez~Garcia, M.~T{\'o}rtola and J.W.F.~Valle, \emph{{Physics implications of a combined analysis of COHERENT CsI and LAr data}}, \href{https://doi.org/10.1007/JHEP04(2023)035}{\emph{JHEP} {\bfseries 04} (2023) 035} [\href{https://arxiv.org/abs/2211.11905}{{\ttfamily 2211.11905}}].

\bibitem{Coloma:2023ixt}
P.~Coloma, M.C.~Gonzalez-Garcia, M.~Maltoni, J.P.~Pinheiro and S.~Urrea, \emph{{Global constraints on non-standard neutrino interactions with quarks and electrons}}, \href{https://doi.org/10.1007/JHEP08(2023)032}{\emph{JHEP} {\bfseries 08} (2023) 032} [\href{https://arxiv.org/abs/2305.07698}{{\ttfamily 2305.07698}}].

\bibitem{Khan:2019cvi}
A.N.~Khan and W.~Rodejohann, \emph{{New physics from COHERENT data with an improved quenching factor}}, \href{https://doi.org/10.1103/PhysRevD.100.113003}{\emph{Phys. Rev. D} {\bfseries 100} (2019) 113003} [\href{https://arxiv.org/abs/1907.12444}{{\ttfamily 1907.12444}}].

\bibitem{Coloma:2022avw}
P.~Coloma, I.~Esteban, M.C.~Gonzalez-Garcia, L.~Larizgoitia, F.~Monrabal and S.~Palomares-Ruiz, \emph{{Bounds on new physics with data of the Dresden-II reactor experiment and COHERENT}}, \href{https://doi.org/10.1007/JHEP05(2022)037}{\emph{JHEP} {\bfseries 05} (2022) 037} [\href{https://arxiv.org/abs/2202.10829}{{\ttfamily 2202.10829}}].

\bibitem{Giunti:2019aiy}
C.~Giunti and T.~Lasserre, \emph{{eV-scale Sterile Neutrinos}}, \href{https://doi.org/10.1146/annurev-nucl-101918-023755}{\emph{Ann. Rev. Nucl. Part. Sci.} {\bfseries 69} (2019) 163} [\href{https://arxiv.org/abs/1901.08330}{{\ttfamily 1901.08330}}].

\bibitem{PhysRevD.86.013004}
A.J.~Anderson, J.M.~Conrad, E.~Figueroa-Feliciano, C.~Ignarra, G.~Karagiorgi, K.~Scholberg et~al., \emph{Measuring active-to-sterile neutrino oscillations with neutral current coherent neutrino-nucleus scattering}, \href{https://doi.org/10.1103/PhysRevD.86.013004}{\emph{Phys. Rev. D} {\bfseries 86} (2012) 013004}.

\bibitem{PhysRevD.96.063013}
T.S.~Kosmas, D.K.~Papoulias, M.~T\'ortola and J.W.F.~Valle, \emph{Probing light sterile neutrino signatures at reactor and spallation neutron source neutrino experiments}, \href{https://doi.org/10.1103/PhysRevD.96.063013}{\emph{Phys. Rev. D} {\bfseries 96} (2017) 063013}.

\bibitem{PhysRevD.101.075051}
C.~Blanco, D.~Hooper and P.~Machado, \emph{Constraining sterile neutrino interpretations of the lsnd and miniboone anomalies with coherent neutrino scattering experiments}, \href{https://doi.org/10.1103/PhysRevD.101.075051}{\emph{Phys. Rev. D} {\bfseries 101} (2020) 075051}.

\bibitem{Giunti:2022aea}
C.~Giunti, J.~Gruszko, B.~Jones, L.~Kaufman, D.~Parno and A.~Pocar, \emph{{Report of the Topical Group on Neutrino Properties for Snowmass 2021}},  \href{https://arxiv.org/abs/2209.03340}{{\ttfamily 2209.03340}}.

\bibitem{AtzoriCorona:2022qrf}
M.~Atzori~Corona, M.~Cadeddu, N.~Cargioli, F.~Dordei, C.~Giunti, Y.F.~Li et~al., \emph{{Impact of the Dresden-II and COHERENT neutrino scattering data on neutrino electromagnetic properties and electroweak physics}}, \href{https://doi.org/10.1007/JHEP09(2022)164}{\emph{JHEP} {\bfseries 09} (2022) 164} [\href{https://arxiv.org/abs/2205.09484}{{\ttfamily 2205.09484}}].

\bibitem{AtzoriCorona:2022moj}
M.~Atzori~Corona, M.~Cadeddu, N.~Cargioli, F.~Dordei, C.~Giunti, Y.F.~Li et~al., \emph{{Probing light mediators and (g {\ensuremath{-}} 2)$_{\mu}$ through detection of coherent elastic neutrino nucleus scattering at COHERENT}}, \href{https://doi.org/10.1007/JHEP05(2022)109}{\emph{JHEP} {\bfseries 05} (2022) 109} [\href{https://arxiv.org/abs/2202.11002}{{\ttfamily 2202.11002}}].

\bibitem{Cadeddu:2020nbr}
M.~Cadeddu, N.~Cargioli, F.~Dordei, C.~Giunti, Y.F.~Li, E.~Picciau et~al., \emph{{Constraints on light vector mediators through coherent elastic neutrino nucleus scattering data from COHERENT}}, \href{https://doi.org/10.1007/JHEP01(2021)116}{\emph{JHEP} {\bfseries 01} (2021) 116} [\href{https://arxiv.org/abs/2008.05022}{{\ttfamily 2008.05022}}].

\bibitem{bechteler_faissner_yogeshwar_seyfarth_1984}
H.~Bechteler, H.~Faissner, R.~Yogeshwar and H.~Seyfarth, \emph{The spectrum of $\gamma$ radiation emitted in the frj-1 (merlin) reactor core and moderator region}, .

\bibitem{Park:2017prx}
H.~Park, \emph{{Detecting Dark Photons with Reactor Neutrino Experiments}}, \href{https://doi.org/10.1103/PhysRevLett.119.081801}{\emph{Phys. Rev. Lett.} {\bfseries 119} (2017) 081801} [\href{https://arxiv.org/abs/1705.02470}{{\ttfamily 1705.02470}}].

\bibitem{Ge:2017mcq}
S.-F.~Ge and I.M.~Shoemaker, \emph{{Constraining Photon Portal Dark Matter with Texono and Coherent Data}}, \href{https://doi.org/10.1007/JHEP11(2018)066}{\emph{JHEP} {\bfseries 11} (2018) 066} [\href{https://arxiv.org/abs/1710.10889}{{\ttfamily 1710.10889}}].

\bibitem{AristizabalSierra:2020rom}
D.~Aristizabal~Sierra, V.~De~Romeri, L.J.~Flores and D.K.~Papoulias, \emph{{Axionlike particles searches in reactor experiments}}, \href{https://doi.org/10.1007/JHEP03(2021)294}{\emph{JHEP} {\bfseries 03} (2021) 294} [\href{https://arxiv.org/abs/2010.15712}{{\ttfamily 2010.15712}}].

\bibitem{Deniverville:2020rbv}
P.~Deniverville, H.-S.~Lee and Y.-M.~Lee, \emph{{New searches at reactor experiments based on the dark axion portal}}, \href{https://doi.org/10.1103/PhysRevD.103.075006}{\emph{Phys. Rev. D} {\bfseries 103} (2021) 075006} [\href{https://arxiv.org/abs/2011.03276}{{\ttfamily 2011.03276}}].

\bibitem{KunioKaneta_2016}
{Kunio Kaneta, Hye-Sung Lee, Seokhoon Yun}, \emph{{Portal Connecting Dark Photons and Axions}}, \href{https://doi.org/10.1103/PhysRevLett.118.101802}{\emph{Phys. Rev. Lett} {\bfseries 118} (2017) 101802} [\href{https://arxiv.org/abs/1611.01466}{{\ttfamily 1611.01466}}].

\bibitem{Bechteler_1984}
{H. Bechteler et al.}, \emph{{The spectrum of $\gamma$ radiation emitted in the FRJ-1 (Merlin) reactor core and moderator region}}, {\emph{Inst. fuer Kernphysik} (1984) }.

\bibitem{Aristizabal_2020}
{D. Aristizabal Sierra, V. De Romeri, L. J. Flores, D.K. Papoulias}, \emph{{Axionlike particles searches in reactor experiments}},  \href{https://arxiv.org/abs/2010.15712}{{\ttfamily 2010.15712}}.

\bibitem{HyangKyu_2017}
{HyangKyu Park}, \emph{{Detecting Dark Photon with Reactor Neutrino Experiments}}, \href{https://doi.org/10.1103/PhysRevLett.119.081801}{\emph{Phys. Rev. Lett.} {\bfseries 119} (2017) 081801} [\href{https://arxiv.org/abs/1705.02470}{{\ttfamily 1705.02470}}].

\bibitem{Patrick_2020}
{Patrick deNiverville, Hye-Sung Lee, Young-Min Lee}, \emph{{New searches at the reactor experiments based on the dark axion portal}}, \href{https://doi.org/10.1103/PhysRevD.103.075006}{\emph{Phys. Rev. D} {\bfseries 103} (2021) 075006} [\href{https://arxiv.org/abs/2011.03276v2}{{\ttfamily 2011.03276v2}}].

\bibitem{Shao-Feng_2017}
{Shao-Feng Ge, Ian M. Shoemaker}, \emph{{Constraining Photon Portal Dark Matter with Texono and Coherent Data}}, \href{https://doi.org/10.1007/JHEP11%282018%29066}{\emph{JHEP} {\bfseries 11} (2018) 066} [\href{https://arxiv.org/abs/1710.10889}{{\ttfamily 1710.10889}}].

\bibitem{XCOM}
M.~Berger, J.~Hubbell, S.~Seltzer, J.~Chang, J.~Coursey, S.~R. et~al., \emph{XCOM: Photon Cross Sections Database}.
\newblock \url{https://dx.doi.org/10.18434/T48G6X}.

\bibitem{deNiverville:2019xsx}
P.~deNiverville and H.-S.~Lee, \emph{{Implications of the dark axion portal for SHiP and FASER and the advantages of monophoton signals}}, \href{https://doi.org/10.1103/PhysRevD.100.055017}{\emph{Phys. Rev. D} {\bfseries 100} (2019) 055017} [\href{https://arxiv.org/abs/1904.13061}{{\ttfamily 1904.13061}}].

\bibitem{Essig:2013vha}
R.~Essig, J.~Mardon, M.~Papucci, T.~Volansky and Y.-M.~Zhong, \emph{{Constraining Light Dark Matter with Low-Energy $e^+e^-$ Colliders}}, \href{https://doi.org/10.1007/JHEP11(2013)167}{\emph{JHEP} {\bfseries 11} (2013) 167} [\href{https://arxiv.org/abs/1309.5084}{{\ttfamily 1309.5084}}].

\bibitem{Arias:2020tzl}
P.~Arias, A.~Arza, J.~Jaeckel and D.~Vargas-Arancibia, \emph{{Hidden Photon Dark Matter Interacting via Axion-like Particles}}, \href{https://doi.org/10.1088/1475-7516/2021/05/070}{\emph{JCAP} {\bfseries 05} (2021) 070} [\href{https://arxiv.org/abs/2007.12585}{{\ttfamily 2007.12585}}].

\bibitem{BaBar:2008aby}
{\scshape BaBar} collaboration, \emph{{Search for Invisible Decays of a Light Scalar in Radiative Transitions $\upsilon_{3S} \to \gamma$ A0}},  in \emph{{34th International Conference on High Energy Physics}}, 7, 2008 [\href{https://arxiv.org/abs/0808.0017}{{\ttfamily 0808.0017}}].

\end{thebibliography}\endgroup

\end{document}